# Superexchange interactions in AgMF$_4$ (M = Co, Ni, Cu) polymorphs

Mateusz A. Domański[1]*, Wojciech Grochala[1]*

[1]Centre of New Technologies, University of Warsaw, S. Banacha 2C, 02-097 Warsaw, Poland

m.domanski@cent.uw.edu.pl, w.grochala@cent.uw.edu.pl

Keywords: theoretical modelling, density functional theory, silver fluorides, magnetic ordering, superexchange

### Abstract

Magnetic properties of silver(II) compounds have been of interest in recent years. In covalent compounds, the main mechanism of interaction between paramagnetic sites is the superexchange *via* the connecting ligand. To date, little is known of magnetic interactions between Ag(II) cations and other paramagnetic centres. It is because only a few compounds bearing Ag(II) cation and other paramagnetic transition metal cation are known experimentally. Recently the high-pressure synthesis of ternary silver(II) fluorides with 3d metal cations AgMF$_4$ (M = Co, Ni, Cu) was predicted to be feasible. Here, we investigate the magnetic properties of these compounds in their diverse polymorphic forms. Using well-established computational methods we predict superexchange pathways in AgMF$_4$, evaluate coupling constants and calculate the impact of Ag(II) presence on superexchange between the other cations. The results indicate that the low-pressure form of AgCuF$_4$, the only composed of stacked layers as the parent AgF$_2$, would hold mainly Ag-Ag and Cu-Cu superexchange interactions. Upon compression, or with the nickel(II) cation, the Ag-M interactions in AgMF$_4$ intensify, which is emphasized by an increase of Ag-M superexchange coupling constants and Ag-F-M angles. All the strongest Ag-M superexchange pathways are quasi-linear, leading to the formation of antiferromagnetic chains along the crystallographic directions. The impact of Ag(II) on M-M superexchange turns out to be moderate, due to factors connected to the crystal structure.

### 1. Introduction

The quest for materials exhibiting large magnetic superexchange (SE) and semiconducting properties abides, especially in the field of rapidly advancing antiferromagnetic spintronics. Such materials are in demand due to their high-frequency response and insensitivity to data-damaging perturbations.[1] However, designing new materials with strong antiferromagnetic coupling is not trivial. The magnetic coupling strength drops sharply with increasing the distance between magnetic centres (due to weakening of orbital overlap) but also with the departure from linear bonding, the angle bending causing the wrong phases of orbitals to overlap.[2] These factors influence the most important feature of antiferromagnetic semiconductors *i.e.* SE interaction between ligand p orbitals and magnetic centre d orbitals, as in the case of oxocuprates[3]. Both these issues are influenced by the structure of the material itself and may be engineered to some extent e.g. by using cations with differing radii (like in series of 1D antiferromagnets bearing silver(II)[4]) or by enforcing the epitaxial growth of single layers[5].

But what are other pathways towards large magnetic SE constant, *J*? Apart from structural changes, commonly the two main ways of enhancing the (anti-)ferromagnetic SE interaction are utilized. The first one called radical-bridged SE consists of placing a free radical placed between interacting magnetic centres.[6] These centres may be connected by paramagnetic (often organic) linkers with a single unpaired electron, which strengthens the direct interactions between metal-based centres. Such significantly localized radical improves *J* through the mechanism of SE mediating the magnetic-centres wavefunctions or through the hopping mechanism.[7]

However, the existence of stable and localized radicals is not necessary, and a formally closed-shell ligand may also do a similar job. The alternative route to increase *J* is by so-called redox non-innocence of ligands, which may be considered as a halfway to the above-mentioned situation of a radical. Here, a redox-active metal centre interacts with the bridging ligand through a strong p-d or s-d (in case of a hydride[8]) hybridization resulting in a partial spin transfer to the ligand. It appears that such small and non-localized magnetic moment induced on non-metal can greatly affect the whole system magnetic properties, mediating (even the long-range) interactions between localized spins.[9] In particular, the ferromagnetic semiconductors are one example of such strongly p-d hybridized systems. They are well described with theory proposed by Dietl et al.,[9] and find applications in modern spintronics. In certain cases, ferroelasticity may appear.[10,11] However, this way also has its limits – in the extreme situation, such a spin transfer may push the system on a verge of stability, eventually when a half electron will transfer to each second unit a whole system may collapse due to phonon instability[12].

The peculiar and attractive materials in this context are fluorides, which have been recently investigated in the context of spintronics. These materials can exhibit extremely considerable anisotropy of spin-electronic properties like $MnF_3$[13] or $[AgF][BF_4]$[14]. The latter is particularly important low-dimensional $d(z^2)^1$ fluoride system with ½ spin exhibiting large exchange anisotropies within antiferromagnetic ground state.[14] Divalent silver is known to be very strong spin polarizing agent, known to polarize even such hard anions like fluoride, where this ligand holds an uncompensated spin density, what was shown on DFT+U level[15].

Therefore, we considered ternary fluoride stoichiometries which may host strongly coupled $AgF_2$ sheets. Among $Ag^{II}M_xF_y$, many transition metal systems have been prepared so far[16], but the great majority of these compounds have closed-shell or low-spin cations. Therefore, we consider alloying silver(II) fluoride with high-spin 3d metal (M = Co, Ni, Cu) fluorides in order to examine their magnetic properties, and check what the influence of ½ spin Ag(II) on M-M magnetic interactions would be. Naturally, we are also interested in the strength of Ag(II)-M magnetic interactions, which have not yet been studied for any among dozens of Ag(II)-TM systems known.[16] In the recent work it was shown, that phases with $AgMF_4$ stoichiometry (M = Co, Ni, Cu) can be synthesized under high pressure conditions, and should be metastable after decompression[17]. Here, we focus on analysis of magnetic interactions inside these yet unknown compounds of $AgMF_4$ stoichiometry, with particular attention to magnetic SE and influence of Ag(II) spin polarizer on M-M interactions.

## 2. Methods

This theoretical study is based on periodic electronic-structure calculations carried out with VASP 5.4.4 software using PAW method[18,19]. We used the potentials set recommended by VASP with a 520 eV energy cut-off. The energy was calculated using collinearly spin-polarized DFT method using GGA functional PBEsol, *i.e.* solid-revised Perdew, Burke and Ernzerhof correlation-exchange functional[20]. The on-site electronic correlation was included with Coulomb and exchange repulsion terms with +U Dudarev's approach[21,22]. In this approach only effective U is taken into account, thus we used $U_{eff}$ equal 4 eV (for Ag, Ni), 5 eV (Co) and 8 (Cu). These values were earlier used and validated in respective systems with +II oxidation state[17,23–26]. We have also cross-checked the applied DFT+U parameters with other DFT approaches, namely SCAN meta-GGA functional and HSE06 hybrid functional, obtaining qualitatively the same results therefore they are not provided in this work. Geometry optimizations were done on DFT+U with fine parameters of 0.024 Å$^{-1}$ wave vector spacing and conjugate-gradient algorithm relaxation with convergence thresholds of 10$^{-7}$ eV (electronic) and 10$^{-5}$ (ionic step).

In the present work, magnetic SE coupling constants were calculated using the broken symmetry method.[30–33] In this method the Heisenberg magnetic interaction between spins are described with broken symmetry high-spin states. Using DFT based methods the magnetic interaction energy may be calculated with the classic Ising model Hamiltonian considering only collinear magnetic moments for any spin value.[32] It was shown that using the broken-symmetry method eigenvalues of the Ising model Hamiltonian correspond to those obtained from Heisenberg Hamiltonian.[30,32] The Ising model Hamiltonian was used in the form $H = -\frac{1}{2} J_{\langle ij \rangle} \sum_{i,j} S_{z,i} S_{z,j}$ (with $J_{\langle ij \rangle} = J_{\langle ji \rangle}$), where $J_{\langle ij \rangle}$ is antiferromagnetic coupling constant between the closest neighbouring magnetic centres $i$ and $j$ when $J_{ij} < 0$. Then, by assuming that all the interactions are additive, all magnetic interactions constants $J_{ij}$ can be extracted solving system of equations derived from different spin states.[30] In case of inconsistencies in a linear system of equations, the solutions were calculated using the least-square method[34]. To include interactions over longer distances, the single-point of different spin states were calculated in 2 x 2 x 2 supercells. Using this method we have obtained coupling constants for the binary fluorides: $CuF_2$, $NiF_2$ and $AgF_2$ (discussed in the main text). The values for $CuF_2$ and $NiF_2$ agree well with ones found in the literature[27–29]. VESTA[35] software was used for the visualization of structures.

### 3. Results

In the present work, we investigate the magnetic properties of the proposed ternary fluorides which contain the spin-polarizing Ag(II) cation. Using the well-established computational methods we want to illustrate the influence of Ag(II) cation on magnetic SE interaction inside the proposed ternary fluorides, which were proven recently to show stability under high pressures[17]. To do so, (i) first we briefly describe the structure of ternary fluorides with formula $AgMF_4$, (ii) next we analyse and calculate the magnitude of magnetic interactions of these compounds and (iii) finally we investigate whether the presence of Ag(II) influence the strength of spin-spin interactions.

#### 3.1. Crystal structure of ternary silver(II) fluorides and SE pathways

Since ternary silver(II) fluorides with 3d metal cations prefer monoclinic distorted $AgF_2$ type structures[17], the parent silver(II) fluoride is a reference essential for a purpose of comparison with the ternary fluorides. In the experimentally known orthorhombic $AgF_2$[36,37] (**Figure 1a**) we distinguish three different magnetic interaction paths named $J_{2D}$, $J_{i1}$ and $J_{i2}$ within distance of about 4.0 Å, following the work of Kurzydłowski et al.[23] (**Figure 1b**). Despite all three interactions are essentially two-dimensional, the $J_{2D}$ is reserved for the strongest interaction within covalently bound $AgF_2$ layers within $ac$ plane. $J_{i1}$ and $J_{i2}$ indicate interactions within $bc$ and ab planes, respectively (*i.e.* interlayer interactions). The ground state of orthorhombic $AgF_2$ holds strong antiferromagnetic interactions within *ac* plane which reveals in highly negative $J_{2D}$ value equal –60.5 meV on DFT+U level (this study). Both $J_{i1}$ and $J_{i2}$ are over one order of magnitude smaller and positive [37]. Values for $J_{2D}$ calculated here on different levels of theory correspond well to the theoretical (–71 meV for SCAN functional[23], –56 meV for DFT+U[38]) as well as experimental (–70 meV[38]) values found in the literature.

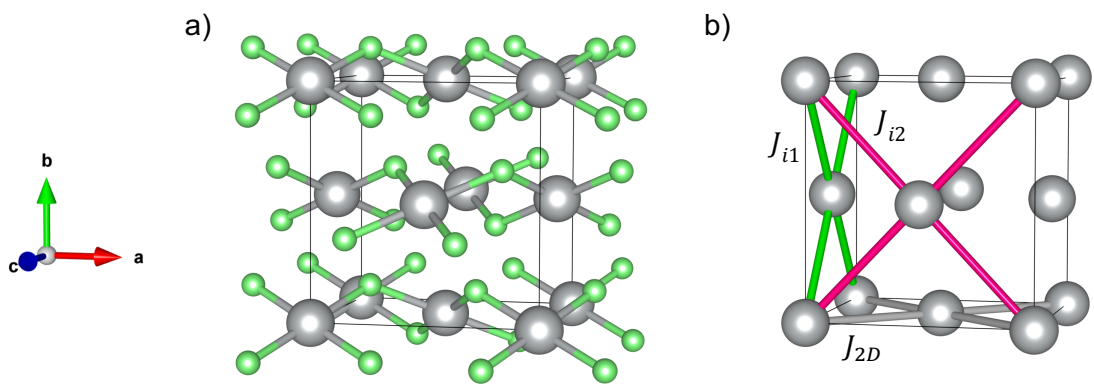

**Figure 1.** (a) Crystal structure of orthorhombic $AgF_2$ at ambient pressure, and (b) magnetic interactions paths between constituting Ag(II) sites. Grey colour for silver atoms, green for fluorine.

Structures of ternary fluorides having 3d cations, proposed in our previous work[17], differ in three major aspects with respect to parent $AgF_2$. The first one is the presence of 3d transition metal cation M(II), which implies differentiating two general types of magnetic interactions – mixed magnetic coupling constants $J^{mix}$ between Ag(II) and M(II), and magnetic constants $J^M$ between two M(II) sites (**Figure 2a**). The second difference is that considered ternary fluorides $AgMF_4$ structures are monoclinically distorted with chosen β > 90°. Due to this distortion, the *ac* face of a monoclinic unit cell is stretched, causing that the original $J_{i2}$ interaction (originally parallel to *ab* plane) transforms from two-dimensional into two distinct linear interactions, which are labelled as $J_3^{mix}$ and $J_4^{mix}$ in **Figure 2b**. This distortion is present in every found $AgMF_4$ ternary fluoride and manifests itself in two different values for Ag-M distances parallel to *ac* plane (distances for $J_3^{mix}$ and $J_4^{mix}$ in **Table 1**). The third difference is in ternary fluorides there are two distinct types of Ag-F-M bridges along the shortest Ag-M distance parallel to *ab* planes (corresponding to $J_{i1}$ parallel to *bc* plane in the parent $AgF_2$). For example, in the $AgNiF_4$ LP structure for an Ag-Ni distance of 3.715 Å there is one Ag-F-Ni bridge with 137.7° angle with short bonds, and the second with a 120.1° angle with weaker bonds (*cf.* **Table S3**). Thus, these two types of bridges are split here into two distinct interactions (*i.e.* to interaction paths $J_1^{mix}$ and $J_2^{mix}$, see **Figure 2b** and **Figure 2c**), despite the same distance between interacting sites.

Moreover, the primary connectivity pattern changes from the layered structure present in $AgF_2$ to the spatially bonded framework, except for the $AgCuF_4$ LP structure which preserves a separated corrugated layer of each difluoride. This change in connectivity implies the greater significance of Ag-M magnetic interaction that these within one-metal layers. This structural difference, however, does not require a different model of magnetic interactions, as the positions of transition metal cations do not change between all presented here structures of ternary fluorides, and they all represent monoclinic distorted $AgF_2$ structure type.

Overall, we considered all interactions within a distance of about 6.0 Å (slightly varying with the formula, see **Table S1**-**Table S5**), resulting in **13** different coupling constants included in the model Hamiltonian. In the model, we recognize: **(1)** Ag(II) sheets coupled with $J_{2D}^{Ag}$; **(2,3,4)**

Ag(II) chains coupled along subsequent lattice vectors *i.e.* $J_y^{Ag}$, $J_z^{Ag}$, $J_x^{Ag}$; **(5,6,7,8)** the same for M(II) with coupling constants $J_{2D}^M$, $J_y^M$, $J_z^M$, $J_x^M$; **(9)** Ag(II)-M(II) chains along [110] or [-110] direction coupled with $J_1^{mix}$ along Ag-F-M bridge with a more obtuse angle and shorter bonds; **(10)** Ag(II)-M(II) chains along [110] or [-110] direction coupled with $J_2^{mix}$ along Ag-F-M bridge with a smaller angle and weaker bonds[39]; **(11)** Ag(II)-M(II) chains along [101] direction coupled with $J_3^{mix}$; **(12)** Ag(II)-M(II) chains [-101] direction coupled with $J_4^{mix}$; **(13)** Ag(II)-M(II) sheets within (-101) plane coupled with $J_5^{mix}$ without a direct SE route.

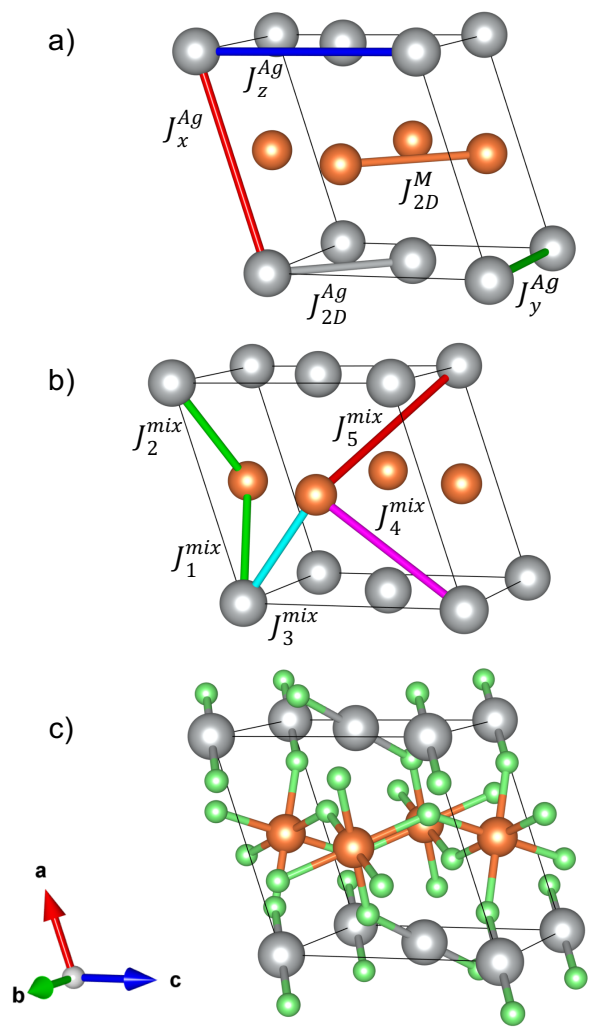

**Figure 2.** Paths of magnetic interactions in the monoclinic distorted structure of AgF$_2$ type, characteristic for every ternary silver(II) fluoride considered in this work. (a) Pure elemental interactions in AgMF$_4$, three interactions along crystallographic axes for M-M interactions are omitted since they are analogous to Ag-Ag ones. (b) Mixed cation interactions in AgMF$_4$, (c) crystal structure typical for AgMF$_4$. Grey colour for silver atoms, orange for M (any considered 3d metal) atoms, green for fluorine atoms (omitted in (a) and (b) for clarity).

### 3.2. Magnetic interactions in AgMF$_4$ systems

After determination of the key SE pathways, we have calculated energies of the possible magnetic configurations of each structure and constructed a system of equations applying the Ising Hamiltonian for each of the structures (provided in points 2. and 3. in SI). Since each of the considered systems of equations was overdetermined, we applied the least-square statistical method to obtain the solutions.

Systems that have been successfully analysed are AgCuF$_4$ and AgNiF$_4$ for both low-pressure polymorphs and high-pressure ones after their decompression to standard pressure (*i.e.* following reoptimization of the structure at p = 0 GPa). In the case of any AgCoF$_4$ system, the instability of magnetic configuration combined with the proximity of intrinsic redox reaction caused that we could not obtain reliable results applying Ising Hamiltonian, nor biquadratic exchange or four-spin ring interactions. The lack of reliable solution manifested as large errors (over 0.1 eV, *cf.* with results from point 3 in SI) between energy values obtained with DFT and energy calculated using Ising Hamiltonian, or modified Hamiltonian including the mentioned interactions. This is why we focus exclusively on the cases of M = Ni and Cu.

**Table 1.** Magnetic SE constants calculated for each ternary silver(II) fluoride system, as well as reference AgF$_2$, provided in meV, and metal-metal separations at the corresponding SE pathways. Distances (d) are provided in angstroms, and coupling constants (J) in meV.

|  | AgF$_2$ | | | AgCuF$_4$ | | | | AgNiF$_4$ | | | | AgCoF$_4$ | |
|---|---|---|---|---|---|---|---|---|---|---|---|---|---|
|  | LP | | | LP | | HP | | LP | | HP | | LP | |
| Interaction | d | J | Interaction | d | J | d | J | d | J | d | J | d | J[c] |
| $J_{2D}$ | 3.743 | −60.5 | $J_{2D}^{Ag}$ | 3.616 | −37.4 | 3.731 | 0.8 | 3.715 | −6.3 | 3.633 | −26.8 | 3.740 | − |
|  |  |  | $J_y^{Ag}$ | 4.764 | −4.0 | 5.293 | −1.7 | 4.804 | −0.7 | 4.909 | −2.1 | 4.797 | − |
|  |  |  | $J_z^{Ag}$ | 5.946 | 2.6 | 5.260 | −0.1 | 5.669 | 0.9 | 5.358 | −0.9 | 5.738 | − |
|  |  |  | $J_x^{Ag}$ | 5.440 | −0.9 | 5.064 | 0.3 | 5.615 | −2.0 | 5.382 | 0.6 | 5.715 | − |
|  |  |  | $J_{2D}^{M}$ | 3.616 | −18.9 | 3.731 | 0.2 | 3.715 | −4.7 | 3.633 | −3.8 | 3.740 | − |
|  |  |  | $J_y^{M}$ | 4.764 | −0.1 | 5.293 | 0.6 | 4.804 | −0.1 | 4.909 | 0.0 | 4.797 | − |
|  |  |  | $J_z^{M}$ | 5.946 | −1.5 | 5.260 | −0.8 | 5.669 | −0.4 | 5.358 | 1.0 | 5.738 | − |
|  |  |  | $J_x^{M}$ | 5.440 | −0.1 | 5.064 | 0.4 | 5.615 | −0.2 | 5.382 | 0.5 | 5.715 | − |
| $J_{i1}$[a] | 3.851 | 5.6 | $J_1^{mix}$ | 3.810 | 4.8 | 3.662 | −45.9 | 3.695 | −32.8 | 3.642 | −33.3 | 3.731 | − |
|  |  |  | $J_2^{mix}$ | 3.810 | 0.0 | 3.662 | 1.3 | 3.695 | −6.9 | 3.642 | 1.9 | 3.731 | − |
| $J_{i2}$[b] | 3.997 | 2.4 | $J_3^{mix}$ | 3.453 | 3.3 | 3.529 | −16.4 | 3.354 | 6.1 | 3.406 | −7.3 | 3.402 | − |
|  |  |  | $J_4^{mix}$ | 4.533 | −0.4 | 3.768 | −0.7 | 4.537 | −1.7 | 4.152 | −1.6 | 4.607 | − |
|  |  |  | $J_5^{mix}$ | 5.884 | −2.2 | 6.497 | −0.5 | 5.859 | −2.4 | 5.975 | −1.6 | 5.859 | − |

[a] interactions $J_1^{mix}$ and $J_2^{mix}$ are identical in AgF$_2$ structure, and equivalent to $J_{i1}$
[b] interactions $J_3^{mix}$ and $J_4^{mix}$ are identical in AgF$_2$ structure, and equivalent to $J_{i2}$
[c] not determined as explained in the text

Magnetic exchange in AgCuF$_4$ LP resembles the situation in parent AgF$_2$ since the strongest interactions are calculated to be the intralayer ones ($J_{2D}^{Ag}$ and $J_{2D}^{Cu}$), with weak ferromagnetic interactions between the layers ($J_1^{mix}$, $J_3^{mix}$). The key change in comparison with AgF$_2$ is that the result of inserting [CuF$_2$] layers is relative flattening of [CuF$_2$] layers (140.8° in AgCuF$_4$ LP, while 131.7° in the pure CuF$_2$) and buckling of [AgF$_2$] sheets (decrease of angle from 129.0° down to 122.2°, cf. **Table S1**), since Cu(II) has much smaller ionic radius than Ag(II). Along with increased buckling in [AgF$_2$] sheets, the intralayer antiferromagnetic exchange decreases for $J_{2D}^{Ag}$ (–60.5 meV for AgF$_2$ vs –37.4 meV for AgCuF$_4$ LP) in accordance with Goodenough-Kanamori-Anderson (GKA) rules[2,40,41]. In both cations, d-holes are majorly occupying d$_{x^2-y^2}$ orbitals as usually in strongly coupled antiferromagnetic layers (**Error! Reference source not found.a**). The obtained spin density projections have been deposited to NoMaD database (https://nomad-lab.eu/).

The core features of the AgCuF$_4$ HP structure are preserved after decompression, but the topology of SE interactions differs dramatically. The groundstate of this structure is ferrimagnetic, with Ag(II) sites having opposite spins to Cu(II) sites, resulting in a small net magnetic moment of about +0.022 µB due to different moments on both cations. In AgCuF$_4$ HP no layers can be separated, what is further confirmed by weak interactions along *bc* planes, $J_{2D}^{Ag}$ and $J_{2D}^{Cu}$. In AgCuF$_4$ HP the shortest interactions and the strongest Ag-F-Cu bridges are $J_1^{mix}$ along directions [110] (or [-110], these directions are alternating along *c* vector), and $J_3^{mix}$ along [10-1]. Holes on both Ag(II) and Cu(II) sites still sit mainly on d$_{x^2-y^2}$ orbitals as in the LP structure (**Error! Reference source not found.**b).

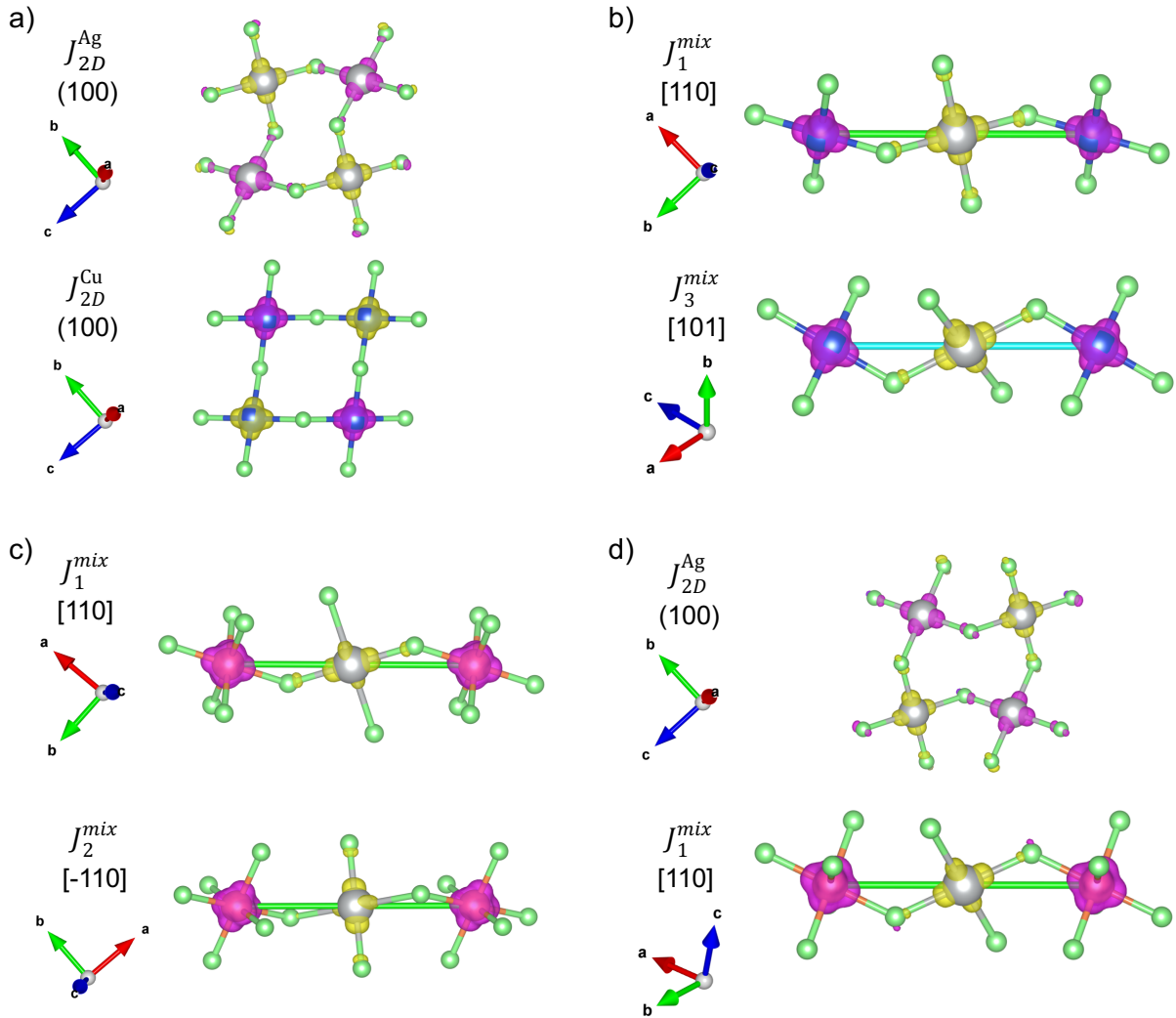

**Figure 3.** The strongest and the second-strongest magnetic coupling SE pathways in AgCuF$_4$ LP (a) and HP (b), as well in AgNiF$_4$ LP (c) and HP (d). Spin-densities are projected on isosurface with a value 0.030 e/Å$^3$, with pink for spin-down and yellow for spin-up densities. The spin-densities are projected for the ground state magnetic configuration for each system. Grey colour for silver, green for fluorine, blue for copper, and orange for nickel atoms.

The interactions along chains dominate also in the AgNiF$_4$ LP structure. Similarly, the strongest are Ag-Ni interactions oriented along directions [110] (or [-110]), namely $J_1^{mix}$. In **Error! Reference source not found.**c the projected spin-density suggests that in this case d-hole in Ag(II) majorly consists of d$_{z^2}$ orbital, enhancing the interaction along $J_1^{mix}$ path. Here, the interactions along the path $J_2^{mix}$ are relatively strong, resulting in the fact that constant $J_2^{mix}$ is the second-largest among all that have been calculated. Unlike in AgCuF$_4$ HP, here $J_3^{mix}$ is positive thus having a ferromagnetic character, because of double Ag-F-Ni bridge having the angle of about 103° (**Table S3**), as usually values of about 90° favour ferromagnetic exchange according to GKA rules.

In the last determined system *i.e.* AgNiF$_4$ HP structure, Ag(II) cations tend to interact within *bc* plane, which is revealed by an increase of $J_{2D}^{Ag}$ from –6.3 meV in LP structure up to –26.8 meV in HP structure, which is about half of the value found for parent AgF$_2$. This structural transition is linked with axial elongation of Ag(II) coordination octahedra and location of spin densities majorly on d$_{x^2-y^2}$ orbitals, just like in parent AgF$_2$ or AgCuF$_4$ LP. Here, however, still the Ag-Ni interaction $J_1^{mix}$ is the strongest one like in AgNiF$_4$ LP structure, suggesting that it may play a crucial role in the predicted stability of the AgNiF$_4$ compound.

### 3.3. Ag(II) influence on M-M SE coupling

Our key aim was to scrutinize, whether the presence of Ag(II) cation impacts magnetic SE between the neighbouring 3d metal cations inside the structures of the proposed ternary fluorides. Therefore, we have conducted a theoretical experiment of substitution paramagnetic Ag(II) for diamagnetic cation with similar charge and very similar ionic radius, Cd(II). In order to investigate the influence of electronic properties of Ag(II) only, we have not reoptimized the structures featuring Cd(II). Next, we recalculated the SE $J_{2D}^M$ as well as other of $J^M$ type, and compared them in **Table 2**.

Ag(II) spin polarizer may influence M-M spin interactions in two basic ways: (i) by inducing the redox non-innocence of ligands, resulting in the increase of the magnetization on fluorides within the [MF$_2$] layers, or (ii) through two consecutive SE paths leading *via* Ag(II) cations such as M'-Ag-M" (*e.g.* through $J_1^{mix}$, $J_2^{mix}$ and $J_3^{mix}$ in **Figure 2b** and **Error! Reference source not found.**), resembling a free radical-bridged next-nearest-neighbour superexchange. (iii) However, one may anticipate a suppression of the orthogonal M-M SE when major interactions are of the mixed Ag-M type, as *e.g.* $J_{2D}^M$ is orthogonal to $J_1^{mix}$ (**Figure 2**). In such case, since Ag(II) active orbitals are oriented along M-Ag-M paths their impact on M-M SE pathways is very small or even weakening. (iv) Also, the influence of Ag(II) would be small, if both Ag(II) and M(II) are significantly separated and interactions between Ag(II) and M(II) are weak.

**Table 2.** Magnetic exchange constants calculated for each ternary silver(II) fluoride system with silver cations $J^M(Ag)$ and after substitution with cadmium cations $J^M(Ag)$, and the ratio between them.

| System | $J^M(Ag)$ / meV | | | | $J^M(Cd)$ / meV | | | | $J^M(Ag)/J^M(Cd)$ | | | |
|---|---|---|---|---|---|---|---|---|---|---|---|---|
| | $J_{2D}^M$ | $J_y^M$ | $J_z^M$ | $J_x^M$ | $J_{2D}^M$ | $J_y^M$ | $J_z^M$ | $J_x^M$ | $J_{2D}^M$ | $J_y^M$ | $J_z^M$ | $J_x^M$ |
| AgCuF$_4$ LP | -18.9 | -0.1 | -1.5 | -0.1 | -17.8 | -0.1 | -0.9 | 0.0 | 106% | 99% | 162% | 2084% |
| AgCuF$_4$ HP | 0.2 | 0.6 | -0.8 | 0.4 | 0.6 | -0.1 | -0.1 | -0.2 | 38% | -841% | 628% | -219% |
| AgNiF$_4$ LP | -4.7 | -0.1 | -0.4 | -0.2 | -5.1 | 0.0 | -0.1 | -0.2 | 91% | -626% | 319% | 87% |
| AgNiF$_4$ HP | -3.9 | 0.0 | 1.0 | 0.5 | -5.1 | 0.0 | -0.1 | -0.1 | 75% | -29% | -753% | -453% |

The results presented in **Table 2** indicate a moderate impact of Ag(II) on the shortest M-M SE interaction, $J_{2D}^M$, ranging between 38% and 106% of the value typical of Cd-substituted compounds. In AgCuF$_4$ LP structure, probably the enhancement by effect (i) dominates but is suppressed by the effect of substantial separation of the layers (iv). In all other structures, the mixed Ag-M SE interactions predominate which is revealed by large $J^{mix}$ constants given in **Table 1**, thus the effect (iii) is operational and the shortest interactions $J_{2D}^M$ are weakened. On the other hand, the influence of Ag(II) on longer and weaker M-M SE interactions $J_y^M$, $J_z^M$ and $J_x^M$ is generally much larger (up to over 2000%), but of course minor in terms of absolute changes, since these interactions are weak anyway. In these cases, probably the mechanism of double SE (ii) or radical-bridging mechanism (i) is essential, and may even cause a change of character of interaction, *i.e.* from ferromagnetic to antiferromagnetic or *vice versa*, as manifested by the minus sign at percent values listed in **Table 2**.

## 4. Conclusions

In this work, the magnetic properties of ternary silver(II) fluorides with nickel and copper were scrutinized. Using the DFT+U computational methods we have predicted SE pathways in ternary silver(II) fluorides, evaluated coupling constants and calculated the impact of Ag(II) presence on SE between the other cations. The model of magnetic interactions described by Ising Hamiltonian requires 13 different coupling constants to obtain a satisfactorily converged solution. The results indicate that the low-pressure form of AgCuF$_4$, the only composed of stacked layers as the parent AgF$_2$, holds mainly $J_{2D}^{Ag}$ and $J_{2D}^{Cu}$ interactions, which are relatively weakened due to different sizes of Ag(II) and Cu(II) cations causing a geometry strain. Upon compression in the AgCuF$_4$ HP, as well as in both LP and HP structures consisting of nickel(II) cation, the linear Ag-M antiferromagnetic interactions are dominating, what is emphasized by substantially negative SE coupling constants and obtuse Ag-F-M angles. The most pronounced antiferromagnetic Ag-Cu SE constant is predicted to be substantial, about –45.9 meV, while the Ag-Ni one is about –33.3 meV. All the strongest Ag-M SE interactions are close-to-linear, leading to the formation of antiferromagnetic chains along the crystallographic directions. Other SE interactions are minor, except for the AgNiF$_4$ HP structure, where $J_{2D}^{Ag}$ is –26.8 meV, still half of the value for parent AgF$_2$. Moreover, the influence of Ag(II) on M-M SE is rather moderate, due to limiting factors such as structurally unfavourable SE pathway, or excessively strong competing Ag-M interactions. The presence of Ag(II) strongly influences, however, most of weak SE coupling constants present in these systems, as compared to the case of diamagnetic Cd(II) replacing Ag(II). It remains to be seen whether Ag(II) could exert strong effects in molecular rather than solid state systems.

### Supporting information

The supporting information contains additional structural data regarding ligand bridges, details of the obtained magnetic configurations and constructed sets of equations with detailed results. The SI is available online on the publisher's website.

### Acknowledgements

This work was supported by Polish National Science Center (NCN) within Beethoven project (2016/23/G/ST5/04320). The research was carried out using supercomputers of Interdisciplinary Centre for Mathematical and Computational Modelling (ICM), University of Warsaw, under grants ADVANCE++ and SAPPHIRE, number GA76-19 and GA83-34.

### References and footnotes

[1] T. Jungwirth, X. Marti, P. Wadley, J. Wunderlich, *Nat. Nanotechnol.* 2016, *11*, 231–241.
[2] P. W. Anderson, *Phys. Rev.* 1950, *79*, 350–356.
[3] M. R. Norman, *Rep. Prog. Phys.* 2016, *79*, 074502.
[4] Z. Mazej, D. Kurzydłowski, W. Grochala, in *Photonic Electron. Prop. Fluoride Mater.* (Eds.: A. Tressaud, K. Poeppelmeier), Elsevier, Boston, 2016, pp. 231–260.
[5] A. Grzelak, H. Su, X. Yang, D. Kurzydłowski, J. Lorenzana, W. Grochala, *Phys. Rev. Mater.* 2020, *4*, 084405.
[6] C. Hua, J. A. DeGayner, T. D. Harris, *Inorg. Chem.* 2019, *58*, 7044–7053.

[7]  C. Lambert, G. Nöll, J. Schelter, *Nat. Mater.* 2002, *1*, 69–73.
[8]  T. Jaroń, W. Grochala, R. Hoffmann, *J. Mol. Model.* 2007, *13*, 769–774.
[9]  T. Dietl, H. Ohno, F. Matsukura, J. Cibert, D. Ferrand, *Science* 2000, *287*, 1019–1022.
[10] X. Xu, Y. Ma, T. Zhang, C. Lei, B. Huang, Y. Dai, *Nanoscale Horiz.* 2020, *5*, 1386–1393.
[11] I. Sánchez-Movellán, J. Moreno-Ceballos, P. García-Fernández, J. A. Aramburu, M. Moreno, *Chem. – Eur. J.* 2021, *n/a*, DOI 10.1002/chem.202101865.
[12] W. Grochala, R. Hoffmann, P. P. Edwards, *Chem. – Eur. J.* 2003, *9*, 575–587.
[13] Y. Jiao, F. Ma, C. Zhang, J. Bell, S. Sanvito, A. Du, *Phys. Rev. Lett.* 2017, *119*, 016403.
[14] D. Kurzydłowski, W. Grochala, *Phys. Rev. B* 2017, *96*, 155140.
[15] Z. Mazej, E. Goreshnik, Z. Jagličić, B. Gaweł, W. Łasocha, D. Grzybowska, T. Jaroń, D. Kurzydłowski, P. Malinowski, W. Koźmiński, J. Szydłowska, P. Leszczyński, W. Grochala, *CrystEngComm* 2009, *11*, 1702–1710.
[16] W. Grochala, R. Hoffmann, *Angew. Chem. Int. Ed.* 2001, *40*, 2742–2781.
[17] M. A. Domański, M. Derzsi, W. Grochala, *RSC Adv.* 2021, *11*, 25801–25810.
[18] G. Kresse, J. Furthmüller, *Phys. Rev. B* 1996, *54*, 11169–11186.
[19] G. Kresse, D. Joubert, *Phys. Rev. B* 1999, *59*, 1758–1775.
[20] J. P. Perdew, A. Ruzsinszky, G. I. Csonka, O. A. Vydrov, G. E. Scuseria, L. A. Constantin, X. Zhou, K. Burke, *Phys. Rev. Lett.* 2008, *100*, 136406.
[21] A. I. Liechtenstein, V. I. Anisimov, J. Zaanen, *Phys. Rev. B* 1995, *52*, R5467–R5470.
[22] S. L. Dudarev, G. A. Botton, S. Y. Savrasov, C. J. Humphreys, A. P. Sutton, *Phys. Rev. B* 1998, *57*, 1505–1509.
[23] D. Kurzydłowski, M. Derzsi, P. Barone, A. Grzelak, V. Struzhkin, J. Lorenzana, W. Grochala, *Chem. Commun.* 2018, *54*, 10252–10255.
[24] J. A. Barreda-Argüeso, S. López-Moreno, M. N. Sanz-Ortiz, F. Aguado, R. Valiente, J. González, F. Rodríguez, A. H. Romero, A. Muñoz, L. Nataf, F. Baudelet, *Phys. Rev. B* 2013, *88*, 214108.
[25] M. Cococcioni, S. de Gironcoli, *Phys. Rev. B* 2005, *71*, 035105.
[26] C. Miller, A. S. Botana, *Phys. Rev. B* 2020, *101*, 195116.
[27] P. Reinhardt, I. de P. R. Moreira, C. de Graaf, R. Dovesi, F. Illas, *Chem. Phys. Lett.* 2000, *319*, 625–630.
[28] M. T. Hutchings, M. F. Thorpe, R. J. Birgeneau, P. A. Fleury, H. J. Guggenheim, *Phys. Rev. B* 1970, *2*, 1362–1373.
[29] I. de P. R. Moreira, R. Dovesi, C. Roetti, V. R. Saunders, R. Orlando, *Phys. Rev. B* 2000, *62*, 7816–7823.
[30] F. Illas, I. P. R. Moreira, C. de Graaf, V. Barone, *Theor. Chem. Acc.* 2000, *104*, 265–272.
[31] D. Dai, M.-H. Whangbo, *J. Chem. Phys.* 2001, *114*, 2887–2893.
[32] D. Dai, M.-H. Whangbo, *J. Chem. Phys.* 2002, *118*, 29–39.
[33] M.-H. Whangbo, H.-J. Koo, D. Dai, *J. Solid State Chem.* 2003, *176*, 417–481.
[34] W. H. Press, S. A. Teukolsky, B. P. Flannery, W. T. Vetterling, *Numerical Recipes in FORTRAN 77: Volume 1, Volume 1 of Fortran Numerical Recipes: The Art of Scientific Computing*, Cambridge University Press, 1992.
[35] K. Momma, F. Izumi, *J. Appl. Crystallogr.* 2011, *44*, 1272–1276.
[36] P. Fischer, D. Schwarzenbach, H. M. Rietveld, *J. Phys. Chem. Solids* 1971, *32*, 543–550.
[37] P. Fischer, G. Roult, D. Schwarzenbach, *J. Phys. Chem. Solids* 1971, *32*, 1641–1647.
[38] J. Gawraczyński, D. Kurzydłowski, R. A. Ewings, S. Bandaru, W. Gadomski, Z. Mazej, G. Ruani, I. Bergenti, T. Jaroń, A. Ozarowski, S. Hill, P. J. Leszczyński, K. Tokár, M. Derzsi, P. Barone, K. Wohlfeld, J. Lorenzana, W. Grochala, *Proc. Natl. Acad. Sci.* 2019, *116*, 1495–1500.
[39] J(1)mix and J(2)mix lie within the same plane and have the same Ag-M distances but are distinguished due to different superexchange paths as explained before., n.d.
[40] J. B. Goodenough, *Phys. Rev.* 1960, *120*, 67–83.
[41] J. Kanamori, *J. Phys. Chem. Solids* 1959, *10*, 87–98.

# SUPPORTING INFORMATION

## 1. Additional structural information

**Table S1.** Key geometric parameters in AgCuF$_4$ LP structure.

| Interaction | d(M'-M") / Å | M'-F-M" angle | d(M'-F) / Å | d(M"-F) / Å |
|---|---|---|---|---|
| $J_{2D}^{Ag}$ | 3.616 | 122.2° | 2.054 (Ag) | 2.077 (Ag) |
| $J_y^{Ag}$ | 4.764 | - | - | - |
| $J_z^{Ag}$ | 5.946 | - | - | - |
| $J_x^{Ag}$ | 5.440 | - | - | - |
| $J_{2D}^{Cu}$ | 3.616 | 140.8° | 1.917 (Cu) | 1.921 (Cu) |
| $J_y^{Cu}$ | 4.764 | - | - | - |
| $J_z^{Cu}$ | 5.946 | - | - | - |
| $J_x^{Cu}$ | 5.440 | - | - | - |
| $J_1^{mix}$ | 3.810 | 127.8° | 2.188 (Cu) | 2.054 (Ag) |
| $J_2^{mix}$ | 3.810 | 115.4° | 1.917 (Cu) | 2.570 (Ag) |
| $J_3^{mix}$ | 3.453 | 108.1° | 2.188 (Cu) | 2.077 (Ag) |
|  |  | 99.5° | 1.921 (Cu) | 2.570 (Ag) |
| $J_4^{mix}$ | 4.533 | - | - | - |
| $J_5^{mix}$ | 5.884 | - | - | - |

**Table S2.** Key geometric parameters in AgCuF$_4$ HP structure.

| Interaction | d(M'-M") / Å | M'-F-M" angle | d(M'-F) / Å | d(M"-F) / Å |
|---|---|---|---|---|
| $J_{2D}^{Ag}$ | 3.731 | 106.4° | 2.064 (Ag) | 2.579 (Ag) |
| $J_y^{Ag}$ | 5.293 | - | - | - |
| $J_z^{Ag}$ | 5.260 | - | - | - |
| $J_x^{Ag}$ | 5.064 | - | - | - |
| $J_{2D}^{Cu}$ | 3.731 | 124.3° | 1.925 (Cu) | 2.291 (Cu) |
| $J_y^{Cu}$ | 5.293 | - | - | - |
| $J_z^{Cu}$ | 5.260 | - | - | - |
| $J_x^{Cu}$ | 5.064 | - | - | - |
| $J_1^{mix}$ | 3.662 | 136.7° | 1.875 (Cu) | 2.064 (Ag) |
| $J_2^{mix}$ | 3.662 | 112.4° | 2.291 (Cu) | 2.115 (Ag) |
| $J_3^{mix}$ | 3.529 | 121.7° (1 path only) | 1.925 (Cu) | 2.115 (Ag) |
| $J_4^{mix}$ | 3.768 | 114.6° | 1.875 (Cu) | 2.579 (Ag) |
| $J_5^{mix}$ | 6.497 | - | - | - |

**Table S3.** Key geometric parameters in AgNiF$_4$ LP structure.

| Interaction | d(M'-M") / Å | M'-F-M" angle | d(M'-F) / Å | d(M"-F) / Å |
|---|---|---|---|---|
| $J_{2D}^{Ag}$ | 3.709 | 116.5° | 2.028 (Ag) | 2.332 (Ag) |
| $J_y^{Ag}$ | 4.805 | - | - | - |
| $J_z^{Ag}$ | 5.650 | - | - | - |
| $J_x^{Ag}$ | 5.624 | - | - | - |
| $J_{2D}^{Ni}$ | 3.709 | 135.8° | 1.995 (Ni) | 2.008 (Ni) |
| $J_y^{Ni}$ | 4.805 | - | - | - |
| $J_z^{Ni}$ | 5.650 | - | - | - |
| $J_x^{Ni}$ | 5.624 | - | - | - |
| $J_1^{mix}$ | 3.699 | 137.7° | 1.941 (Ni) | 2.025 (Ag) |
| $J_2^{mix}$ | 3.699 | 120.0° | 2.008 (Ni) | 2.260 (Ag) |
| $J_3^{mix}$ | 3.350 | 102.9° | 1.941 (Ni) | 2.332 (Ag) |
|  |  | 103.7° | 1.995 (Ni) | 2.260 (Ag) |
| $J_4^{mix}$ | 4.534 | - | - | - |
| $J_5^{mix}$ | 5.858 | - | - | - |

**Table S4.** Key geometric parameters in AgNiF$_4$ HP structure.

| Interaction | d(M'-M") / Å | M'-F-M" angle | d(M'-F) / Å | d(M"-F) / Å |
|---|---|---|---|---|
| $J_{2D}^{Ag}$ | 3.633 | 119.5 | 2.059 (Ag) | 2.146 (Ag) |
| $J_y^{Ag}$ | 4.909 | - | - | - |
| $J_z^{Ag}$ | 5.358 | - | - | - |
| $J_x^{Ag}$ | 5.382 | - | - | - |
| $J_{2D}^{Ni}$ | 3.633 | 134.1 | 1.973 (Ni) | 1.974 (Ni) |
| $J_y^{Ni}$ | 4.909 | - | - | - |
| $J_z^{Ni}$ | 5.358 | - | - | - |
| $J_x^{Ni}$ | 5.382 | - | - | - |
| $J_1^{mix}$ | 3.642 | 128.3° | 1.987 (Ni) | 2.059 (Ag) |
| $J_2^{mix}$ | 3.642 | 100.1° | 1.974 (Ni) | 2.627 (Ag) |
| $J_3^{mix}$ | 3.406 | 110.9° | 1.987 (Ni) | 2.146 (Ag) |
|  |  | 94.5° | 1.973 (Ni) | 2.627 (Ag) |
| $J_4^{mix}$ | 4.152 | 122.9° | 1.974 (Ni) | 2.736 (Ag) |
| $J_5^{mix}$ | 5.975 | - | - | - |

**Table S5.** Key geometric parameters in AgCoF$_4$ LP structure.

| Interaction | d(M'-M") / Å | M'-F-M" angle | d(M'-F) / Å | d(M"-F) / Å |
|---|---|---|---|---|
| $J_{2D}^{Ag}$ | 3.740 | 118.0 | 2.030 (Ag) | 2.330 (Ag) |
| $J_y^{Ag}$ | 4.796 | - | - | - |
| $J_z^{Ag}$ | 5.738 | - | - | - |
| $J_x^{Ag}$ | 5.716 | - | - | - |
| $J_{2D}^{Co}$ | 3.740 | 134.6 | 2.025 (Co) | 2.029 (Co) |
| $J_y^{Co}$ | 4.797 | - | - | - |
| $J_z^{Co}$ | 5.738 | - | - | - |
| $J_x^{Co}$ | 5.716 | - | - | - |
| $J_1^{mix}$ | 3.731 | 136.6° | 1.985 (Co) | 2.030 (Ag) |
| $J_2^{mix}$ | 3.731 | 120.5° | 2.025 (Co) | 2.271 (Ag) |
| $J_3^{mix}$ | 3.402 | 103.8°<br>104.5° | 1.985 (Co)<br>2.029 (Co) | 2.330 (Ag)<br>2.271 (Ag) |
| $J_4^{mix}$ | 4.607 | - | - | - |
| $J_5^{mix}$ | 5.881 | - | - | - |

## 2. Obtained magnetic configurations

**Table S6.** Collinear spin value on magnetic centers for each magnetic configuration of AgCuF$_4$ LP structure, which were used for magnetic SE constants calculations.

| 2·S | State | Ag 1 | 2 | 3 | 4 | 5 | 6 | 7 | 8 | 9 | 10 | 11 | 12 | 13 | 14 | 15 | 16 | Cu 1 | 2 | 3 | 4 | 5 | 6 | 7 | 8 | 9 | 10 | 11 | 12 | 13 | 14 | 15 | 16 |
|---|---|---|---|---|---|---|---|---|---|---|---|---|---|---|---|---|---|---|---|---|---|---|---|---|---|---|---|---|---|---|---|---|---|
| 0 | **AFM_1*** | 0.5 | 0.5 | 0.5 | 0.5 | 0.5 | 0.5 | 0.5 | 0.5 | -0.5 | -0.5 | -0.5 | -0.5 | -0.5 | -0.5 | -0.5 | -0.5 | 0.5 | 0.5 | 0.5 | 0.5 | 0.5 | 0.5 | 0.5 | 0.5 | -0.5 | -0.5 | -0.5 | -0.5 | -0.5 | -0.5 | -0.5 | -0.5 |
| 0 | AFM_2 | -0.5 | -0.5 | -0.5 | -0.5 | -0.5 | -0.5 | -0.5 | -0.5 | 0.5 | 0.5 | 0.5 | 0.5 | 0.5 | 0.5 | 0.5 | 0.5 | 0.5 | 0.5 | 0.5 | 0.5 | 0.5 | 0.5 | 0.5 | 0.5 | -0.5 | -0.5 | -0.5 | -0.5 | -0.5 | -0.5 | -0.5 | -0.5 |
| 0 | AFM_3 | -0.5 | -0.5 | -0.5 | 0.5 | -0.5 | -0.5 | -0.5 | 0.5 | -0.5 | -0.5 | -0.5 | 0.5 | -0.5 | -0.5 | 0.5 | 0.5 | -0.5 | -0.5 | 0.5 | 0.5 | 0.5 | 0.5 | 0.5 | 0.5 | 0.5 | 0.5 | 0.5 | 0.5 | -0.5 | -0.5 | -0.5 | -0.5 |
| 0 | AFM_4 | -0.5 | 0.5 | 0.5 | -0.5 | -0.5 | 0.5 | 0.5 | -0.5 | 0.5 | 0.5 | -0.5 | -0.5 | 0.5 | 0.5 | 0.5 | -0.5 | 0.5 | 0.5 | 0.5 | 0.5 | 0.5 | 0.5 | 0.5 | 0.5 | -0.5 | -0.5 | -0.5 | -0.5 | -0.5 | -0.5 | -0.5 | -0.5 |
| 0 | AFM_5 | -0.5 | -0.5 | 0.5 | 0.5 | -0.5 | -0.5 | 0.5 | 0.5 | -0.5 | -0.5 | 0.5 | 0.5 | -0.5 | -0.5 | 0.5 | 0.5 | -0.5 | -0.5 | 0.5 | 0.5 | -0.5 | -0.5 | 0.5 | 0.5 | -0.5 | -0.5 | 0.5 | 0.5 | -0.5 | -0.5 | 0.5 | 0.5 |
| 0 | AFM_6 | -0.5 | -0.5 | -0.5 | -0.5 | 0.5 | 0.5 | 0.5 | 0.5 | -0.5 | -0.5 | -0.5 | -0.5 | 0.5 | 0.5 | 0.5 | 0.5 | -0.5 | -0.5 | 0.5 | 0.5 | -0.5 | -0.5 | 0.5 | 0.5 | -0.5 | -0.5 | 0.5 | 0.5 | -0.5 | -0.5 | 0.5 | 0.5 |
| 0 | AFM_7 | -0.5 | -0.5 | -0.5 | -0.5 | -0.5 | -0.5 | -0.5 | -0.5 | 0.5 | 0.5 | 0.5 | 0.5 | 0.5 | 0.5 | 0.5 | 0.5 | -0.5 | -0.5 | 0.5 | 0.5 | -0.5 | -0.5 | 0.5 | 0.5 | -0.5 | -0.5 | 0.5 | 0.5 | -0.5 | -0.5 | 0.5 | 0.5 |
| 0 | AFM_8 | -0.5 | 0.5 | 0.5 | -0.5 | 0.5 | -0.5 | -0.5 | 0.5 | 0.5 | -0.5 | -0.5 | 0.5 | -0.5 | 0.5 | 0.5 | -0.5 | -0.5 | -0.5 | 0.5 | 0.5 | -0.5 | -0.5 | 0.5 | 0.5 | -0.5 | -0.5 | 0.5 | 0.5 | -0.5 | -0.5 | 0.5 | 0.5 |
| 0 | AFM_9 | -0.5 | 0.5 | 0.5 | 0.5 | -0.5 | -0.5 | -0.5 | 0.5 | -0.5 | 0.5 | 0.5 | 0.5 | -0.5 | -0.5 | 0.5 | 0.5 | -0.5 | -0.5 | -0.5 | 0.5 | 0.5 | 0.5 | -0.5 | -0.5 | -0.5 | 0.5 | 0.5 | 0.5 | -0.5 | -0.5 | 0.5 | 0.5 |
| 0 | AFM_10 | -0.5 | -0.5 | -0.5 | -0.5 | 0.5 | 0.5 | 0.5 | 0.5 | -0.5 | -0.5 | -0.5 | -0.5 | 0.5 | 0.5 | 0.5 | 0.5 | -0.5 | -0.5 | -0.5 | -0.5 | 0.5 | 0.5 | 0.5 | 0.5 | -0.5 | -0.5 | -0.5 | -0.5 | 0.5 | 0.5 | 0.5 | 0.5 |
| 0 | AFM_11 | -0.5 | -0.5 | -0.5 | -0.5 | 0.5 | 0.5 | 0.5 | 0.5 | 0.5 | 0.5 | 0.5 | 0.5 | -0.5 | -0.5 | -0.5 | -0.5 | -0.5 | -0.5 | -0.5 | -0.5 | 0.5 | 0.5 | 0.5 | 0.5 | -0.5 | -0.5 | -0.5 | -0.5 | 0.5 | 0.5 | 0.5 | 0.5 |
| 0 | AFM_12 | 0.5 | 0.5 | 0.5 | 0.5 | -0.5 | -0.5 | -0.5 | -0.5 | 0.5 | 0.5 | 0.5 | 0.5 | -0.5 | -0.5 | -0.5 | -0.5 | -0.5 | -0.5 | -0.5 | -0.5 | 0.5 | 0.5 | 0.5 | 0.5 | -0.5 | -0.5 | -0.5 | -0.5 | -0.5 | -0.5 | -0.5 | -0.5 |
| 0 | AFM_13 | -0.5 | 0.5 | 0.5 | -0.5 | -0.5 | 0.5 | 0.5 | -0.5 | 0.5 | -0.5 | -0.5 | 0.5 | 0.5 | -0.5 | -0.5 | 0.5 | 0.5 | 0.5 | 0.5 | 0.5 | -0.5 | -0.5 | -0.5 | -0.5 | -0.5 | -0.5 | -0.5 | -0.5 | 0.5 | 0.5 | 0.5 | 0.5 |
| 0 | AFM_14 | -0.5 | -0.5 | 0.5 | 0.5 | 0.5 | 0.5 | -0.5 | -0.5 | 0.5 | 0.5 | -0.5 | -0.5 | -0.5 | -0.5 | 0.5 | 0.5 | 0.5 | 0.5 | -0.5 | -0.5 | -0.5 | -0.5 | 0.5 | 0.5 | -0.5 | -0.5 | 0.5 | 0.5 | 0.5 | 0.5 | -0.5 | -0.5 |
| 0 | AFM_15 | -0.5 | -0.5 | 0.5 | 0.5 | 0.5 | 0.5 | -0.5 | -0.5 | -0.5 | -0.5 | 0.5 | 0.5 | 0.5 | 0.5 | -0.5 | -0.5 | 0.5 | 0.5 | -0.5 | -0.5 | 0.5 | 0.5 | -0.5 | -0.5 | -0.5 | -0.5 | 0.5 | 0.5 | -0.5 | -0.5 | 0.5 | 0.5 |
| 0 | AFM_16 | -0.5 | 0.5 | 0.5 | -0.5 | 0.5 | -0.5 | -0.5 | 0.5 | 0.5 | -0.5 | -0.5 | 0.5 | -0.5 | 0.5 | 0.5 | -0.5 | 0.5 | -0.5 | -0.5 | 0.5 | -0.5 | 0.5 | 0.5 | -0.5 | -0.5 | 0.5 | 0.5 | -0.5 | 0.5 | -0.5 | -0.5 | 0.5 |
| 0 | AFM_17 | -0.5 | 0.5 | 0.5 | -0.5 | -0.5 | 0.5 | 0.5 | -0.5 | 0.5 | -0.5 | -0.5 | 0.5 | 0.5 | -0.5 | -0.5 | 0.5 | 0.5 | -0.5 | 0.5 | -0.5 | 0.5 | -0.5 | 0.5 | -0.5 | -0.5 | 0.5 | -0.5 | 0.5 | -0.5 | 0.5 | -0.5 | 0.5 |
| 0 | AFM_18 | -0.5 | -0.5 | 0.5 | 0.5 | -0.5 | -0.5 | 0.5 | 0.5 | 0.5 | 0.5 | -0.5 | -0.5 | 0.5 | 0.5 | -0.5 | -0.5 | 0.5 | 0.5 | -0.5 | -0.5 | 0.5 | 0.5 | -0.5 | -0.5 | 0.5 | 0.5 | -0.5 | -0.5 | 0.5 | 0.5 | -0.5 | -0.5 |
| 16 | FM16_1 | 0.5 | 0.5 | 0.5 | 0.5 | 0.5 | 0.5 | 0.5 | 0.5 | 0.5 | 0.5 | 0.5 | 0.5 | 0.5 | 0.5 | 0.5 | 0.5 | 0.5 | 0.5 | 0.5 | 0.5 | 0.5 | 0.5 | 0.5 | 0.5 | -0.5 | -0.5 | -0.5 | -0.5 | -0.5 | -0.5 | -0.5 | -0.5 |
| 16 | FM16_2 | 0.5 | 0.5 | 0.5 | 0.5 | 0.5 | 0.5 | 0.5 | 0.5 | 0.5 | 0.5 | 0.5 | 0.5 | 0.5 | 0.5 | 0.5 | 0.5 | -0.5 | -0.5 | -0.5 | -0.5 | 0.5 | 0.5 | 0.5 | 0.5 | -0.5 | -0.5 | -0.5 | -0.5 | 0.5 | 0.5 | 0.5 | 0.5 |
| 16 | FM16_3 | 0.5 | 0.5 | 0.5 | 0.5 | 0.5 | 0.5 | 0.5 | 0.5 | 0.5 | 0.5 | 0.5 | 0.5 | 0.5 | 0.5 | 0.5 | 0.5 | -0.5 | -0.5 | 0.5 | 0.5 | -0.5 | -0.5 | 0.5 | 0.5 | -0.5 | -0.5 | 0.5 | 0.5 | -0.5 | -0.5 | 0.5 | 0.5 |
| 16 | FM16_4 | -0.5 | 0.5 | 0.5 | 0.5 | 0.5 | -0.5 | 0.5 | 0.5 | 0.5 | 0.5 | -0.5 | 0.5 | 0.5 | 0.5 | 0.5 | -0.5 | 0.5 | 0.5 | 0.5 | 0.5 | 0.5 | 0.5 | 0.5 | 0.5 | 0.5 | 0.5 | 0.5 | 0.5 | 0.5 | 0.5 | 0.5 | 0.5 |
| 16 | FM16_5 | -0.5 | 0.5 | 0.5 | -0.5 | -0.5 | 0.5 | 0.5 | -0.5 | 0.5 | 0.5 | 0.5 | 0.5 | 0.5 | 0.5 | 0.5 | 0.5 | 0.5 | 0.5 | 0.5 | 0.5 | 0.5 | 0.5 | 0.5 | 0.5 | -0.5 | -0.5 | 0.5 | 0.5 | -0.5 | -0.5 | 0.5 | 0.5 |
| 32 | FM32 | 0.5 | 0.5 | 0.5 | 0.5 | 0.5 | 0.5 | 0.5 | 0.5 | 0.5 | 0.5 | 0.5 | 0.5 | 0.5 | 0.5 | 0.5 | 0.5 | 0.5 | 0.5 | 0.5 | 0.5 | 0.5 | 0.5 | 0.5 | 0.5 | 0.5 | 0.5 | 0.5 | 0.5 | 0.5 | 0.5 | 0.5 | 0.5 |

\* groundstate

**Table S7.** Collinear spin value on magnetic centers for each magnetic configuration of AgCuF$_4$ HP structure, which were used for magnetic SE constants calculations.

| 2·S | State | Ag 1 | 2 | 3 | 4 | 5 | 6 | 7 | 8 | 9 | 10 | 11 | 12 | 13 | 14 | 15 | 16 | Cu 1 | 2 | 3 | 4 | 5 | 6 | 7 | 8 | 9 | 10 | 11 | 12 | 13 | 14 | 15 | 16 |
|---|---|---|---|---|---|---|---|---|---|---|---|---|---|---|---|---|---|---|---|---|---|---|---|---|---|---|---|---|---|---|---|---|---|
| 0 | AFM_1 | 0.5 | 0.5 | 0.5 | 0.5 | 0.5 | 0.5 | 0.5 | 0.5 | -0.5 | -0.5 | -0.5 | -0.5 | -0.5 | -0.5 | -0.5 | -0.5 | 0.5 | 0.5 | 0.5 | 0.5 | 0.5 | 0.5 | 0.5 | 0.5 | -0.5 | -0.5 | -0.5 | -0.5 | -0.5 | -0.5 | -0.5 | -0.5 |
| 0 | AFM_2 | -0.5 | -0.5 | -0.5 | -0.5 | -0.5 | -0.5 | -0.5 | -0.5 | 0.5 | 0.5 | 0.5 | 0.5 | 0.5 | 0.5 | 0.5 | 0.5 | 0.5 | 0.5 | 0.5 | 0.5 | 0.5 | 0.5 | 0.5 | 0.5 | -0.5 | -0.5 | -0.5 | -0.5 | -0.5 | -0.5 | -0.5 | -0.5 |

| 2·S | State | Ag 1 | 2 | 3 | 4 | 5 | 6 | 7 | 8 | 9 | 10 | 11 | 12 | 13 | 14 | 15 | 16 | Ni 1 | 2 | 3 | 4 | 5 | 6 | 7 | 8 | 9 | 10 | 11 | 12 | 13 | 14 | 15 | 16 |
|---|---|---|---|---|---|---|---|---|---|---|---|---|---|---|---|---|---|---|---|---|---|---|---|---|---|---|---|---|---|---|---|---|---|
| 0 | AFM_3 | -0.5 | -0.5 | -0.5 | -0.5 | -0.5 | -0.5 | 0.5 | 0.5 | -0.5 | -0.5 | 0.5 | 0.5 | -0.5 | -0.5 | 0.5 | 0.5 | -0.5 | -0.5 | 0.5 | 0.5 | 0.5 | 0.5 | 0.5 | 0.5 | 0.5 | 0.5 | 0.5 | 0.5 | -0.5 | -0.5 | -0.5 | -0.5 |
| 0 | AFM_4 | -0.5 | 0.5 | 0.5 | -0.5 | -0.5 | 0.5 | 0.5 | -0.5 | -0.5 | 0.5 | 0.5 | -0.5 | -0.5 | 0.5 | 0.5 | -0.5 | 0.5 | 0.5 | 0.5 | 0.5 | 0.5 | 0.5 | 0.5 | 0.5 | -0.5 | -0.5 | -0.5 | -0.5 | -0.5 | -0.5 | -0.5 | -0.5 |
| 0 | AFM_5 | -0.5 | -0.5 | 0.5 | 0.5 | 0.5 | 0.5 | 0.5 | 0.5 | -0.5 | -0.5 | -0.5 | -0.5 | -0.5 | -0.5 | -0.5 | -0.5 | 0.5 | 0.5 | 0.5 | 0.5 | 0.5 | 0.5 | 0.5 | 0.5 | -0.5 | -0.5 | -0.5 | -0.5 | -0.5 | -0.5 | -0.5 | -0.5 |
| 0 | AFM_6 | -0.5 | -0.5 | -0.5 | -0.5 | -0.5 | -0.5 | -0.5 | -0.5 | 0.5 | 0.5 | 0.5 | 0.5 | 0.5 | 0.5 | 0.5 | 0.5 | 0.5 | 0.5 | 0.5 | 0.5 | 0.5 | 0.5 | 0.5 | 0.5 | -0.5 | -0.5 | -0.5 | -0.5 | -0.5 | -0.5 | -0.5 | -0.5 |
| 0 | AFM_7 | -0.5 | -0.5 | -0.5 | -0.5 | 0.5 | 0.5 | 0.5 | 0.5 | -0.5 | -0.5 | -0.5 | -0.5 | 0.5 | 0.5 | 0.5 | 0.5 | -0.5 | -0.5 | 0.5 | 0.5 | -0.5 | -0.5 | 0.5 | 0.5 | -0.5 | -0.5 | 0.5 | 0.5 | -0.5 | -0.5 | 0.5 | 0.5 |
| 0 | AFM_8 | -0.5 | -0.5 | 0.5 | 0.5 | -0.5 | -0.5 | 0.5 | 0.5 | -0.5 | -0.5 | 0.5 | 0.5 | -0.5 | -0.5 | 0.5 | 0.5 | -0.5 | -0.5 | 0.5 | 0.5 | -0.5 | -0.5 | 0.5 | 0.5 | -0.5 | -0.5 | 0.5 | 0.5 | -0.5 | -0.5 | 0.5 | 0.5 |
| 0 | AFM_9 | -0.5 | -0.5 | -0.5 | -0.5 | 0.5 | 0.5 | 0.5 | 0.5 | -0.5 | -0.5 | -0.5 | -0.5 | 0.5 | 0.5 | 0.5 | 0.5 | -0.5 | -0.5 | -0.5 | -0.5 | 0.5 | 0.5 | 0.5 | 0.5 | -0.5 | -0.5 | -0.5 | -0.5 | 0.5 | 0.5 | 0.5 | 0.5 |
| 0 | AFM_10 | -0.5 | -0.5 | -0.5 | -0.5 | -0.5 | -0.5 | -0.5 | -0.5 | 0.5 | 0.5 | 0.5 | 0.5 | 0.5 | 0.5 | 0.5 | 0.5 | -0.5 | -0.5 | -0.5 | -0.5 | -0.5 | -0.5 | -0.5 | -0.5 | 0.5 | 0.5 | 0.5 | 0.5 | 0.5 | 0.5 | 0.5 | 0.5 |
| 0 | AFM_11 | 0.5 | 0.5 | 0.5 | 0.5 | 0.5 | 0.5 | 0.5 | 0.5 | -0.5 | -0.5 | -0.5 | -0.5 | -0.5 | -0.5 | -0.5 | -0.5 | 0.5 | 0.5 | 0.5 | 0.5 | -0.5 | -0.5 | -0.5 | -0.5 | -0.5 | -0.5 | -0.5 | -0.5 | 0.5 | 0.5 | 0.5 | 0.5 |
| 0 | AFM_12 | -0.5 | -0.5 | 0.5 | 0.5 | -0.5 | -0.5 | 0.5 | 0.5 | -0.5 | -0.5 | 0.5 | 0.5 | -0.5 | -0.5 | 0.5 | 0.5 | 0.5 | 0.5 | -0.5 | -0.5 | 0.5 | 0.5 | -0.5 | -0.5 | 0.5 | 0.5 | -0.5 | -0.5 | 0.5 | 0.5 | -0.5 | -0.5 |
| 0 | AFM_13 | 0.5 | 0.5 | 0.5 | 0.5 | -0.5 | -0.5 | -0.5 | -0.5 | 0.5 | 0.5 | 0.5 | 0.5 | -0.5 | -0.5 | -0.5 | -0.5 | -0.5 | -0.5 | 0.5 | 0.5 | 0.5 | 0.5 | -0.5 | -0.5 | -0.5 | -0.5 | 0.5 | 0.5 | 0.5 | 0.5 | -0.5 | -0.5 |
| 0 | AFM_14 | -0.5 | 0.5 | 0.5 | -0.5 | 0.5 | -0.5 | -0.5 | 0.5 | 0.5 | -0.5 | -0.5 | 0.5 | -0.5 | 0.5 | 0.5 | -0.5 | 0.5 | -0.5 | -0.5 | 0.5 | -0.5 | 0.5 | 0.5 | -0.5 | -0.5 | 0.5 | 0.5 | -0.5 | 0.5 | -0.5 | -0.5 | 0.5 |
| 0 | AFM_15 | -0.5 | -0.5 | -0.5 | -0.5 | 0.5 | 0.5 | 0.5 | 0.5 | 0.5 | 0.5 | 0.5 | 0.5 | -0.5 | -0.5 | -0.5 | -0.5 | 0.5 | 0.5 | 0.5 | 0.5 | -0.5 | -0.5 | -0.5 | -0.5 | -0.5 | -0.5 | -0.5 | -0.5 | 0.5 | 0.5 | 0.5 | 0.5 |
| 0 | AFM_16 | -0.5 | -0.5 | -0.5 | -0.5 | 0.5 | 0.5 | 0.5 | 0.5 | -0.5 | -0.5 | -0.5 | -0.5 | 0.5 | 0.5 | 0.5 | 0.5 | 0.5 | 0.5 | 0.5 | 0.5 | -0.5 | -0.5 | -0.5 | -0.5 | 0.5 | 0.5 | 0.5 | 0.5 | -0.5 | -0.5 | -0.5 | -0.5 |
| 0 | AFM_16 | -0.5 | -0.5 | 0.5 | 0.5 | -0.5 | -0.5 | 0.5 | 0.5 | -0.5 | -0.5 | 0.5 | 0.5 | -0.5 | -0.5 | 0.5 | 0.5 | 0.5 | -0.5 | -0.5 | 0.5 | 0.5 | -0.5 | -0.5 | 0.5 | 0.5 | -0.5 | -0.5 | 0.5 | 0.5 | -0.5 | -0.5 | 0.5 |
| 0 | AFM_17 | -0.5 | -0.5 | 0.5 | 0.5 | -0.5 | -0.5 | 0.5 | 0.5 | 0.5 | 0.5 | -0.5 | -0.5 | 0.5 | 0.5 | -0.5 | -0.5 | 0.5 | -0.5 | -0.5 | 0.5 | 0.5 | -0.5 | -0.5 | 0.5 | -0.5 | 0.5 | 0.5 | -0.5 | -0.5 | 0.5 | 0.5 | -0.5 |
| 0 | AFM_18 | -0.5 | -0.5 | 0.5 | 0.5 | -0.5 | -0.5 | 0.5 | 0.5 | -0.5 | -0.5 | 0.5 | 0.5 | -0.5 | -0.5 | 0.5 | 0.5 | 0.5 | 0.5 | -0.5 | -0.5 | 0.5 | 0.5 | -0.5 | -0.5 | 0.5 | 0.5 | -0.5 | -0.5 | 0.5 | 0.5 | -0.5 | -0.5 |
| 0 | **AFM_19*** | **-0.5** | **-0.5** | **-0.5** | **-0.5** | **-0.5** | **-0.5** | **-0.5** | **-0.5** | **-0.5** | **-0.5** | **-0.5** | **-0.5** | **-0.5** | **-0.5** | **-0.5** | **-0.5** | **0.5** | **0.5** | **0.5** | **0.5** | **0.5** | **0.5** | **0.5** | **0.5** | **0.5** | **0.5** | **0.5** | **0.5** | **0.5** | **0.5** | **0.5** | **0.5** |
| 8 | FM8_1 | 0.5 | 0.5 | 0.5 | 0.5 | 0.5 | 0.5 | 0.5 | 0.5 | 0.5 | 0.5 | 0.5 | 0.5 | -0.5 | -0.5 | -0.5 | -0.5 | 0.5 | 0.5 | 0.5 | 0.5 | 0.5 | 0.5 | 0.5 | 0.5 | 0.5 | 0.5 | -0.5 | -0.5 | -0.5 | -0.5 | -0.5 | -0.5 |
| 8 | FM8_2 | 0.5 | 0.5 | 0.5 | 0.5 | 0.5 | 0.5 | 0.5 | 0.5 | 0.5 | 0.5 | 0.5 | 0.5 | -0.5 | -0.5 | -0.5 | -0.5 | -0.5 | -0.5 | -0.5 | -0.5 | -0.5 | -0.5 | 0.5 | 0.5 | 0.5 | 0.5 | 0.5 | 0.5 | 0.5 | 0.5 | 0.5 | 0.5 |
| 8 | FM8_3 | 0.5 | 0.5 | 0.5 | 0.5 | 0.5 | 0.5 | 0.5 | 0.5 | 0.5 | 0.5 | 0.5 | 0.5 | -0.5 | -0.5 | -0.5 | -0.5 | 0.5 | 0.5 | -0.5 | -0.5 | -0.5 | -0.5 | -0.5 | -0.5 | 0.5 | 0.5 | 0.5 | 0.5 | 0.5 | 0.5 | -0.5 | -0.5 |
| 8 | FM8_4 | 0.5 | 0.5 | 0.5 | 0.5 | 0.5 | 0.5 | 0.5 | 0.5 | 0.5 | 0.5 | 0.5 | 0.5 | -0.5 | -0.5 | -0.5 | -0.5 | -0.5 | -0.5 | 0.5 | 0.5 | 0.5 | 0.5 | 0.5 | 0.5 | -0.5 | -0.5 | -0.5 | -0.5 | -0.5 | -0.5 | 0.5 | 0.5 |
| 16 | FM16_1 | 0.5 | 0.5 | 0.5 | 0.5 | 0.5 | 0.5 | 0.5 | 0.5 | 0.5 | 0.5 | 0.5 | 0.5 | 0.5 | 0.5 | 0.5 | 0.5 | 0.5 | 0.5 | 0.5 | 0.5 | 0.5 | 0.5 | 0.5 | 0.5 | -0.5 | -0.5 | -0.5 | -0.5 | -0.5 | -0.5 | -0.5 | -0.5 |
| 16 | FM16_2 | 0.5 | 0.5 | 0.5 | 0.5 | 0.5 | 0.5 | 0.5 | 0.5 | 0.5 | 0.5 | 0.5 | 0.5 | 0.5 | 0.5 | 0.5 | 0.5 | 0.5 | 0.5 | 0.5 | 0.5 | 0.5 | 0.5 | 0.5 | 0.5 | -0.5 | -0.5 | -0.5 | -0.5 | -0.5 | -0.5 | -0.5 | -0.5 |
| 16 | FM16_3 | 0.5 | 0.5 | 0.5 | 0.5 | 0.5 | 0.5 | 0.5 | 0.5 | 0.5 | 0.5 | 0.5 | 0.5 | 0.5 | 0.5 | 0.5 | 0.5 | 0.5 | 0.5 | 0.5 | 0.5 | 0.5 | 0.5 | -0.5 | -0.5 | -0.5 | -0.5 | -0.5 | -0.5 | 0.5 | 0.5 | 0.5 | 0.5 |
| 16 | FM16_4 | 0.5 | 0.5 | 0.5 | 0.5 | 0.5 | 0.5 | 0.5 | 0.5 | 0.5 | 0.5 | 0.5 | 0.5 | 0.5 | 0.5 | 0.5 | 0.5 | -0.5 | -0.5 | 0.5 | 0.5 | -0.5 | -0.5 | 0.5 | 0.5 | -0.5 | -0.5 | 0.5 | 0.5 | -0.5 | -0.5 | 0.5 | 0.5 |
| 16 | FM16_5 | 0.5 | 0.5 | 0.5 | 0.5 | 0.5 | 0.5 | 0.5 | 0.5 | 0.5 | 0.5 | 0.5 | 0.5 | 0.5 | 0.5 | 0.5 | 0.5 | -0.5 | -0.5 | 0.5 | 0.5 | -0.5 | -0.5 | 0.5 | 0.5 | -0.5 | -0.5 | 0.5 | 0.5 | -0.5 | -0.5 | 0.5 | 0.5 |
| 16 | FM16_6 | -0.5 | 0.5 | 0.5 | -0.5 | 0.5 | -0.5 | -0.5 | 0.5 | 0.5 | -0.5 | -0.5 | 0.5 | -0.5 | 0.5 | 0.5 | -0.5 | 0.5 | -0.5 | 0.5 | 0.5 | -0.5 | 0.5 | -0.5 | -0.5 | 0.5 | -0.5 | 0.5 | 0.5 | -0.5 | 0.5 | -0.5 | -0.5 |
| 16 | FM16_7 | -0.5 | 0.5 | 0.5 | -0.5 | 0.5 | -0.5 | -0.5 | 0.5 | 0.5 | -0.5 | -0.5 | 0.5 | -0.5 | 0.5 | 0.5 | -0.5 | 0.5 | 0.5 | 0.5 | 0.5 | 0.5 | 0.5 | 0.5 | 0.5 | -0.5 | -0.5 | -0.5 | -0.5 | -0.5 | -0.5 | -0.5 | -0.5 |
| 32 | FM32 | 0.5 | 0.5 | 0.5 | 0.5 | 0.5 | 0.5 | 0.5 | 0.5 | 0.5 | 0.5 | 0.5 | 0.5 | 0.5 | 0.5 | 0.5 | 0.5 | 0.5 | 0.5 | 0.5 | 0.5 | 0.5 | 0.5 | 0.5 | 0.5 | 0.5 | 0.5 | 0.5 | 0.5 | 0.5 | 0.5 | 0.5 | 0.5 |

* groundstate

**Table S8.** Collinear spin value on magnetic centers for each magnetic configuration of AgNiF$_4$ LP structure, which were used for magnetic SE constants calculations.

| 2·S | State | Ag 1 | 2 | 3 | 4 | 5 | 6 | 7 | 8 | 9 | 10 | 11 | 12 | 13 | 14 | 15 | 16 | Ni 1 | 2 | 3 | 4 | 5 | 6 | 7 | 8 | 9 | 10 | 11 | 12 | 13 | 14 | 15 | 16 |
|---|---|---|---|---|---|---|---|---|---|---|---|---|---|---|---|---|---|---|---|---|---|---|---|---|---|---|---|---|---|---|---|---|---|
| 0 | AFM_1 | -0.5 | -0.5 | -0.5 | -0.5 | 0.5 | 0.5 | 0.5 | 0.5 | -0.5 | -0.5 | -0.5 | -0.5 | 0.5 | 0.5 | 0.5 | 0.5 | -1 | -1 | -1 | -1 | 1 | 1 | 1 | 1 | -1 | -1 | -1 | -1 | 1 | 1 | 1 | 1 |
| 0 | **AFM_2*** | **-0.5** | **-0.5** | **-0.5** | **-0.5** | **-0.5** | **-0.5** | **-0.5** | **-0.5** | **0.5** | **0.5** | **0.5** | **0.5** | **0.5** | **0.5** | **0.5** | **0.5** | **1** | **1** | **1** | **1** | **1** | **1** | **1** | **1** | **-1** | **-1** | **-1** | **-1** | **-1** | **-1** | **-1** | **-1** |
| 0 | AFM_3 | -0.5 | -0.5 | -0.5 | -0.5 | -0.5 | -0.5 | -0.5 | -0.5 | 0.5 | 0.5 | 0.5 | 0.5 | 0.5 | 0.5 | 0.5 | 0.5 | -1 | -1 | -1 | -1 | -1 | -1 | -1 | -1 | 1 | 1 | 1 | 1 | 1 | 1 | 1 | 1 |
| 0 | AFM_4 | 0.5 | 0.5 | 0.5 | 0.5 | -0.5 | -0.5 | -0.5 | -0.5 | -0.5 | -0.5 | -0.5 | -0.5 | 0.5 | 0.5 | 0.5 | 0.5 | -1 | -1 | -1 | -1 | 1 | 1 | 1 | 1 | 1 | 1 | 1 | 1 | -1 | -1 | -1 | -1 |
| 0 | AFM_5 | 0.5 | 0.5 | -0.5 | -0.5 | 0.5 | 0.5 | -0.5 | -0.5 | 0.5 | 0.5 | -0.5 | -0.5 | 0.5 | 0.5 | -0.5 | -0.5 | -1 | -1 | 1 | 1 | -1 | -1 | 1 | 1 | -1 | -1 | 1 | 1 | -1 | -1 | 1 | 1 |
| 0 | AFM_6 | 0.5 | 0.5 | -0.5 | -0.5 | 0.5 | 0.5 | -0.5 | -0.5 | 0.5 | 0.5 | -0.5 | -0.5 | -0.5 | -0.5 | 0.5 | 0.5 | -1 | -1 | 1 | 1 | -1 | -1 | 1 | 1 | -1 | -1 | 1 | 1 | -1 | -1 | 1 | 1 |
| 0 | AFM_7 | 0.5 | 0.5 | -0.5 | -0.5 | 0.5 | 0.5 | -0.5 | -0.5 | 0.5 | 0.5 | -0.5 | -0.5 | 0.5 | 0.5 | -0.5 | -0.5 | -1 | -1 | 1 | 1 | -1 | -1 | 1 | 1 | 1 | 1 | -1 | -1 | 1 | 1 | -1 | -1 |
| 0 | AFM_8 | 0.5 | 0.5 | -0.5 | -0.5 | 0.5 | 0.5 | -0.5 | -0.5 | 0.5 | 0.5 | -0.5 | -0.5 | 0.5 | 0.5 | -0.5 | -0.5 | 1 | 1 | -1 | -1 | 1 | 1 | -1 | -1 | 1 | 1 | -1 | -1 | 1 | 1 | -1 | -1 |
| 0 | AFM_9 | -0.5 | 0.5 | -0.5 | 0.5 | -0.5 | 0.5 | -0.5 | 0.5 | -0.5 | 0.5 | -0.5 | 0.5 | -0.5 | 0.5 | -0.5 | 0.5 | 1 | -1 | 1 | -1 | 1 | -1 | 1 | -1 | 1 | -1 | 1 | -1 | 1 | -1 | 1 | -1 |
| 0 | AFM_10 | -0.5 | 0.5 | 0.5 | -0.5 | 0.5 | -0.5 | -0.5 | 0.5 | 0.5 | -0.5 | -0.5 | 0.5 | -0.5 | 0.5 | 0.5 | -0.5 | 1 | -1 | -1 | 1 | -1 | 1 | 1 | -1 | -1 | 1 | 1 | -1 | 1 | -1 | -1 | 1 |
| 0 | AFM_11 | -0.5 | -0.5 | 0.5 | 0.5 | -0.5 | -0.5 | 0.5 | 0.5 | -0.5 | -0.5 | 0.5 | 0.5 | -0.5 | -0.5 | 0.5 | 0.5 | 1 | -1 | 1 | -1 | 1 | -1 | 1 | -1 | 1 | -1 | 1 | -1 | 1 | -1 | 1 | -1 |
| 0 | AFM_12 | -0.5 | -0.5 | 0.5 | 0.5 | -0.5 | -0.5 | 0.5 | 0.5 | -0.5 | -0.5 | 0.5 | 0.5 | -0.5 | -0.5 | 0.5 | 0.5 | -1 | -1 | -1 | -1 | 1 | 1 | 1 | 1 | -1 | -1 | -1 | -1 | 1 | 1 | 1 | 1 |
| 0 | AFM_13 | 0.5 | -0.5 | 0.5 | -0.5 | 0.5 | -0.5 | 0.5 | -0.5 | 0.5 | -0.5 | 0.5 | -0.5 | 0.5 | -0.5 | 0.5 | -0.5 | -1 | 1 | -1 | 1 | -1 | 1 | -1 | 1 | -1 | 1 | -1 | 1 | -1 | 1 | -1 | 1 |
| 0 | AFM_14 | -0.5 | 0.5 | -0.5 | 0.5 | -0.5 | 0.5 | -0.5 | 0.5 | -0.5 | 0.5 | -0.5 | 0.5 | -0.5 | 0.5 | -0.5 | 0.5 | -1 | 1 | -1 | 1 | -1 | 1 | -1 | 1 | -1 | 1 | -1 | 1 | -1 | 1 | -1 | 1 |
| 4 | FM4 | 0.5 | 0.5 | -0.5 | -0.5 | 0.5 | 0.5 | -0.5 | -0.5 | 0.5 | 0.5 | -0.5 | -0.5 | 0.5 | 0.5 | -0.5 | -0.5 | -1 | -1 | 1 | 1 | -1 | -1 | 1 | 1 | -1 | -1 | 1 | 1 | -1 | -1 | 1 | 1 |
| 8 | FM8 | 0.5 | 0.5 | 0.5 | 0.5 | -0.5 | -0.5 | -0.5 | -0.5 | 0.5 | 0.5 | 0.5 | 0.5 | -0.5 | -0.5 | -0.5 | -0.5 | -1 | -1 | -1 | -1 | 1 | 1 | 1 | 1 | -1 | -1 | -1 | -1 | 1 | 1 | 1 | 1 |
| 10 | FM10 | 0.5 | 0.5 | -0.5 | -0.5 | 0.5 | 0.5 | -0.5 | -0.5 | 0.5 | 0.5 | -0.5 | -0.5 | 0.5 | 0.5 | -0.5 | -0.5 | -1 | -1 | -1 | -1 | -1 | -1 | -1 | -1 | -1 | -1 | -1 | -1 | -1 | -1 | -1 | -1 |
| 16 | FM16_1 | 0.5 | 0.5 | 0.5 | 0.5 | 0.5 | 0.5 | 0.5 | 0.5 | 0.5 | 0.5 | 0.5 | 0.5 | 0.5 | 0.5 | 0.5 | 0.5 | -1 | -1 | -1 | -1 | -1 | -1 | -1 | -1 | 1 | 1 | 1 | 1 | 1 | 1 | 1 | 1 |
| 16 | FM16_2 | 0.5 | 0.5 | 0.5 | 0.5 | 0.5 | 0.5 | 0.5 | 0.5 | 0.5 | 0.5 | 0.5 | 0.5 | 0.5 | 0.5 | 0.5 | 0.5 | -1 | -1 | -1 | -1 | 1 | 1 | 1 | 1 | 1 | 1 | 1 | 1 | -1 | -1 | -1 | -1 |
| 16 | FM16_3 | -0.5 | -0.5 | -0.5 | -0.5 | -0.5 | -0.5 | -0.5 | -0.5 | -0.5 | -0.5 | -0.5 | -0.5 | -0.5 | -0.5 | -0.5 | -0.5 | 1 | 1 | 1 | 1 | 1 | 1 | 1 | 1 | 1 | 1 | 1 | 1 | 1 | 1 | 1 | 1 |
| 24 | FM24 | 0.5 | 0.5 | -0.5 | -0.5 | 0.5 | 0.5 | -0.5 | -0.5 | 0.5 | 0.5 | -0.5 | -0.5 | 0.5 | 0.5 | -0.5 | -0.5 | 1 | 1 | -1 | -1 | 1 | 1 | -1 | -1 | 1 | 1 | -1 | -1 | 1 | 1 | -1 | -1 |
| 32 | FM32 | 0.5 | 0.5 | 0.5 | 0.5 | -0.5 | -0.5 | -0.5 | -0.5 | 0.5 | 0.5 | 0.5 | 0.5 | -0.5 | -0.5 | -0.5 | -0.5 | 1 | 1 | 1 | 1 | 1 | 1 | 1 | 1 | 1 | 1 | 1 | 1 | 1 | 1 | 1 | 1 |
| 48 | FM48 | 0.5 | 0.5 | 0.5 | 0.5 | 0.5 | 0.5 | 0.5 | 0.5 | 0.5 | 0.5 | 0.5 | 0.5 | 0.5 | 0.5 | 0.5 | 0.5 | 1 | 1 | 1 | 1 | 1 | 1 | 1 | 1 | 1 | 1 | 1 | 1 | 1 | 1 | 1 | 1 |

* groundstate

**Table S9.** Collinear spin value on magnetic centers for each magnetic configuration of AgNiF$_4$ HP structure, which were used for magnetic SE constants calculations.

| 2·S | State | Ag 1 | 2 | 3 | 4 | 5 | 6 | 7 | 8 | 9 | 10 | 11 | 12 | 13 | 14 | 15 | 16 | Ni 1 | 2 | 3 | 4 | 5 | 6 | 7 | 8 | 9 | 10 | 11 | 12 | 13 | 14 | 15 | 16 |
|---|---|---|---|---|---|---|---|---|---|---|---|---|---|---|---|---|---|---|---|---|---|---|---|---|---|---|---|---|---|---|---|---|---|
| 0 | AFM_1 | -0.5 | -0.5 | -0.5 | -0.5 | 0.5 | 0.5 | 0.5 | 0.5 | 0.5 | -0.5 | -0.5 | -0.5 | -0.5 | 0.5 | 0.5 | 0.5 | -1 | -1 | -1 | -1 | 1 | 1 | 1 | 1 | 1 | 1 | 1 | 1 | -1 | -1 | -1 | -1 |
| 0 | AFM_2 | 0.5 | 0.5 | 0.5 | 0.5 | 0.5 | 0.5 | 0.5 | 0.5 | -0.5 | -0.5 | -0.5 | -0.5 | -0.5 | -0.5 | -0.5 | -0.5 | 1 | 1 | 1 | 1 | 1 | 1 | 1 | 1 | -1 | -1 | -1 | -1 | -1 | -1 | -1 | -1 |
| **0** | **AFM_3*** | **0.5** | **0.5** | **0.5** | **0.5** | **0.5** | **0.5** | **0.5** | **0.5** | **-0.5** | **-0.5** | **-0.5** | **-0.5** | **-0.5** | **-0.5** | **-0.5** | **-0.5** | **-1** | **-1** | **-1** | **-1** | **-1** | **-1** | **-1** | **-1** | **1** | **1** | **1** | **1** | **1** | **1** | **1** | **1** |
| 0 | AFM_4 | -0.5 | -0.5 | -0.5 | -0.5 | 0.5 | 0.5 | 0.5 | 0.5 | 0.5 | 0.5 | 0.5 | 0.5 | -0.5 | -0.5 | -0.5 | -0.5 | -1 | -1 | -1 | -1 | 1 | 1 | 1 | 1 | -1 | -1 | -1 | -1 | 1 | 1 | 1 | 1 |
| 0 | AFM_5 | 0.5 | 0.5 | -0.5 | -0.5 | 0.5 | 0.5 | -0.5 | -0.5 | 0.5 | 0.5 | -0.5 | -0.5 | 0.5 | 0.5 | -0.5 | 0.5 | 1 | 1 | -1 | -1 | 1 | 1 | -1 | -1 | -1 | -1 | 1 | 1 | -1 | -1 | 1 | 1 |
| 0 | AFM_6 | 0.5 | 0.5 | -0.5 | -0.5 | -0.5 | -0.5 | 0.5 | 0.5 | -0.5 | -0.5 | 0.5 | 0.5 | 0.5 | 0.5 | -0.5 | -0.5 | 1 | 1 | -1 | -1 | 1 | 1 | -1 | -1 | -1 | -1 | 1 | 1 | -1 | -1 | 1 | 1 |
| 0 | AFM_7 | 0.5 | 0.5 | -0.5 | -0.5 | 0.5 | 0.5 | -0.5 | -0.5 | 0.5 | 0.5 | -0.5 | -0.5 | 0.5 | 0.5 | -0.5 | -0.5 | 1 | 1 | -1 | -1 | -1 | -1 | 1 | 1 | 1 | 1 | -1 | -1 | -1 | -1 | 1 | 1 |
| 0 | AFM_8 | 0.5 | 0.5 | 0.5 | 0.5 | -0.5 | -0.5 | -0.5 | -0.5 | 0.5 | 0.5 | 0.5 | 0.5 | -0.5 | -0.5 | -0.5 | -0.5 | -1 | -1 | 1 | 1 | -1 | -1 | 1 | 1 | -1 | -1 | 1 | 1 | -1 | -1 | 1 | 1 |
| 0 | AFM_9 | -0.5 | -0.5 | -0.5 | -0.5 | 0.5 | 0.5 | 0.5 | 0.5 | -0.5 | -0.5 | -0.5 | -0.5 | 0.5 | 0.5 | 0.5 | 0.5 | 1 | -1 | 1 | -1 | 1 | -1 | 1 | -1 | 1 | -1 | 1 | -1 | 1 | -1 | 1 | -1 |
| 0 | AFM_10 | -0.5 | 0.5 | 0.5 | -0.5 | 0.5 | -0.5 | -0.5 | 0.5 | -0.5 | 0.5 | 0.5 | -0.5 | 0.5 | -0.5 | -0.5 | 0.5 | -1 | 1 | -1 | 1 | 1 | -1 | 1 | -1 | 1 | -1 | 1 | -1 | -1 | 1 | -1 | 1 |
| 0 | AFM_11 | -0.5 | 0.5 | -0.5 | 0.5 | -0.5 | 0.5 | -0.5 | 0.5 | 0.5 | -0.5 | 0.5 | -0.5 | 0.5 | -0.5 | 0.5 | -0.5 | -1 | 1 | -1 | 1 | 1 | 1 | 1 | 1 | -1 | -1 | -1 | -1 | 1 | -1 | -1 | -1 |
| 0 | AFM_12 | -0.5 | 0.5 | 0.5 | -0.5 | -0.5 | 0.5 | 0.5 | -0.5 | -0.5 | 0.5 | 0.5 | -0.5 | -0.5 | 0.5 | 0.5 | -0.5 | 1 | 1 | 1 | -1 | -1 | 1 | 1 | -1 | -1 | 1 | 1 | -1 | -1 | 1 | -1 | 1 |
| 4 | FM4 | -0.5 | -0.5 | 0.5 | 0.5 | 0.5 | 0.5 | -0.5 | -0.5 | 0.5 | 0.5 | -0.5 | -0.5 | -0.5 | -0.5 | 0.5 | 0.5 | 1 | 1 | -1 | -1 | 1 | 1 | -1 | -1 | -1 | -1 | 1 | 1 | -1 | -1 | 1 | 1 |
| 8 | FM8 | 0.5 | 0.5 | 0.5 | 0.5 | -0.5 | -0.5 | -0.5 | -0.5 | 0.5 | 0.5 | 0.5 | 0.5 | -0.5 | -0.5 | -0.5 | -0.5 | -1 | -1 | -1 | -1 | 1 | 1 | 1 | 1 | 1 | 1 | 1 | 1 | -1 | -1 | -1 | -1 |
| 16 | FM16_1 | 0.5 | -0.5 | 0.5 | -0.5 | -0.5 | 0.5 | -0.5 | 0.5 | 0.5 | -0.5 | 0.5 | -0.5 | -0.5 | 0.5 | -0.5 | 0.5 | 1 | -1 | -1 | 1 | -1 | -1 | -1 | -1 | -1 | -1 | -1 | -1 | 1 | 1 | -1 | 1 |
| 16 | FM16_2 | 0.5 | 0.5 | 0.5 | 0.5 | 0.5 | 0.5 | 0.5 | 0.5 | 0.5 | 0.5 | 0.5 | 0.5 | 0.5 | 0.5 | 0.5 | 0.5 | -1 | -1 | -1 | -1 | -1 | -1 | -1 | -1 | -1 | -1 | -1 | -1 | -1 | -1 | -1 | -1 |
| 16 | FM16_3 | 0.5 | 0.5 | -0.5 | -0.5 | 0.5 | 0.5 | -0.5 | -0.5 | 0.5 | 0.5 | -0.5 | -0.5 | 0.5 | 0.5 | -0.5 | -0.5 | 1 | 1 | -1 | -1 | 1 | 1 | -1 | -1 | -1 | -1 | 1 | 1 | -1 | -1 | 1 | 1 |
| 16 | FM16_4 | -0.5 | -0.5 | -0.5 | -0.5 | -0.5 | -0.5 | -0.5 | -0.5 | -0.5 | -0.5 | -0.5 | -0.5 | -0.5 | -0.5 | -0.5 | -0.5 | 1 | 1 | 1 | 1 | 1 | 1 | 1 | 1 | 1 | 1 | 1 | 1 | 1 | 1 | 1 | 1 |
| 32 | FM32 | 0.5 | 0.5 | 0.5 | 0.5 | -0.5 | -0.5 | -0.5 | -0.5 | 0.5 | 0.5 | 0.5 | 0.5 | -0.5 | -0.5 | -0.5 | -0.5 | 1 | 1 | 1 | 1 | 1 | 1 | 1 | 1 | 1 | 1 | 1 | 1 | 1 | 1 | 1 | 1 |
| 48 | FM48 | 0.5 | 0.5 | 0.5 | 0.5 | 0.5 | 0.5 | 0.5 | 0.5 | 0.5 | 0.5 | 0.5 | 0.5 | 0.5 | 0.5 | 0.5 | 0.5 | 1 | 1 | 1 | 1 | 1 | 1 | 1 | 1 | 1 | 1 | 1 | 1 | 1 | 1 | 1 | 1 |

* groundstate

## 3. Sets of equations

**Table S10.** Coefficients of each type $J$ in every calculated magnetic confuguration of AgCuF$_4$ LP structure, calculated using formulas derived from Heisenberg Hamiltonian. Energies obtained from DFT calculations ($E_{DFT}$), Ising Hamiltonian ($E_H$) and differences between them ($\Delta E$, solution's error for a given state) are provided in eV. All values correspond to 2 x 2 x 2 supercell (Z=16).

| 2·S | State | $E_0$ | $J_{2D}^{Ag}$ | $J_y^{Ag}$ | $J_z^{Ag}$ | $J_x^{Ag}$ | $J_{2D}^{M}$ | $J_y^{M}$ | $J_z^{M}$ | $J_x^{M}$ | $J_1^{mix}$ | $J_2^{mix}$ | $J_3^{mix}$ | $J_4^{mix}$ | $J_5^{mix}$ | $E_{DFT}$ | $E_H$ | $\Delta E$ |
|---|---|---|---|---|---|---|---|---|---|---|---|---|---|---|---|---|---|---|
| **0** | **AFM_1*** | **1** | **8** | **-4** | **-4** | **-4** | **8** | **-4** | **-4** | **-4** | **-8** | **-8** | **8** | **8** | **16** | **-329.8155** | **-329.8135** | **0.0019** |
| 0 | AFM_2 | 1 | 8 | -4 | -4 | -4 | 8 | -4 | -4 | -4 | 8 | 8 | -8 | -8 | -16 | -329.7124 | -329.7144 | -0.0019 |
| 0 | AFM_3 | 1 | 0 | 2 | -4 | -2 | 4 | -2 | -4 | 2 | 0 | 2 | -2 | -2 | 4 | -329.3870 | -329.3801 | 0.0069 |
| 0 | AFM_4 | 1 | 0 | 4 | 4 | -4 | 8 | -4 | -4 | -4 | 0 | 0 | 0 | 0 | 0 | -329.4745 | -329.4766 | -0.0020 |
| 0 | AFM_5 | 1 | 8 | -4 | -4 | 4 | 8 | -4 | -4 | -4 | 0 | 0 | 0 | 0 | 0 | -329.4010 | -329.4045 | -0.0035 |
| 0 | AFM_6 | 1 | -8 | -4 | -4 | 4 | 8 | -4 | -4 | -4 | 0 | 0 | 0 | 0 | 0 | -329.0229 | -329.0224 | 0.0004 |
| 0 | AFM_7 | 1 | 0 | 4 | -4 | -4 | 0 | 4 | -4 | -4 | 0 | 0 | -8 | -8 | 16 | -329.6150 | -329.6137 | 0.0014 |
| 0 | AFM_8 | 1 | -8 | -4 | -4 | 4 | 0 | 4 | -4 | -4 | 0 | 0 | 0 | 0 | 0 | -329.3255 | -329.3263 | -0.0008 |
| 0 | AFM_9 | 1 | 8 | -4 | -4 | -4 | 0 | 4 | -4 | -4 | 0 | 0 | 0 | 0 | 0 | -329.1942 | -329.1958 | -0.0016 |
| 0 | AFM_10 | 1 | 0 | 4 | 4 | -4 | 0 | 4 | -4 | -4 | 0 | 0 | 0 | 0 | 0 | -328.8726 | -328.8713 | 0.0013 |
| 0 | AFM_11 | 1 | 8 | -4 | -4 | -4 | 8 | -4 | -4 | 4 | 0 | 0 | 0 | 0 | 0 | -329.4679 | -329.4695 | -0.0016 |
| 0 | AFM_12 | 1 | 8 | -4 | -4 | -4 | 8 | -4 | -4 | 4 | 0 | 0 | 0 | 0 | 0 | -329.4365 | -329.4329 | 0.0036 |
| 0 | AFM_13 | 1 | 0 | 2 | -4 | -2 | 4 | -2 | -4 | -2 | -2 | 0 | 0 | 0 | 8 | -329.4746 | -329.4771 | -0.0024 |
| 0 | AFM_14 | 1 | 0 | 4 | 4 | -4 | 8 | -4 | -4 | 4 | 0 | 0 | 0 | 0 | 0 | -329.7673 | -329.7714 | -0.0041 |
| 0 | AFM_15 | 1 | 8 | -4 | -4 | 4 | 8 | -4 | -4 | 4 | 0 | 0 | 0 | 0 | 0 | -329.1721 | -329.1732 | -0.0012 |
| 0 | AFM_16 | 1 | -8 | -4 | -4 | 4 | 8 | -4 | -4 | -4 | 0 | 0 | 0 | 0 | 0 | -329.3355 | -329.3414 | -0.0059 |
| 0 | AFM_17 | 1 | 0 | 4 | 4 | 4 | 0 | 4 | 4 | 4 | 0 | 0 | 8 | -8 | -16 | -329.3353 | -329.3388 | -0.0035 |
| 0 | AFM_18 | 1 | 0 | 3 | 3 | 3 | 0 | 3 | 3 | 3 | -2 | 2 | -6 | 6 | 12 | -329.2877 | -329.2892 | -0.0016 |

| 2·S | State | $E_0$ | $J_{2D}^{Ag}$ | $J_y^{Ag}$ | $J_z^{Ag}$ | $J_x^{Ag}$ | $J_{2D}^{M}$ | $J_y^{M}$ | $J_z^{M}$ | $J_x^{M}$ | $J_1^{mix}$ | $J_2^{mix}$ | $J_3^{mix}$ | $J_4^{mix}$ | $J_5^{mix}$ | $E_{DFT}$ | $E_H$ | $\Delta E$ |
|---|---|---|---|---|---|---|---|---|---|---|---|---|---|---|---|---|---|---|
| 16 | FM16_1 | 1 | 0 | 4 | -4 | -4 | 0 | 4 | -4 | -4 | 0 | 0 | 8 | 8 | -16 | -329.1645 | -329.1658 | -0.0013 |
| 16 | FM16_2 | 1 | -8 | -4 | -4 | -4 | -8 | -4 | -4 | -4 | 8 | 8 | 8 | 8 | 16 | -329.0155 | -329.0155 | -0.0001 |
| 16 | FM16_3 | 1 | 0 | -4 | -4 | 0 | 8 | -4 | -4 | -4 | -4 | -4 | 4 | 4 | 8 | -329.0255 | -329.0272 | -0.0017 |
| 16 | FM16_4 | 1 | 0 | -4 | -4 | 0 | 8 | -4 | -4 | -4 | 4 | 4 | -4 | -4 | -8 | -329.2319 | -329.2232 | 0.0087 |
| 16 | FM16_5 | 1 | 0 | -4 | -4 | 0 | -8 | -4 | -4 | 4 | 0 | 0 | 0 | 0 | 0 | -329.3365 | -329.3221 | 0.0144 |
| 32 | FM32 | 1 | 0 | -4 | -4 | 0 | 8 | -4 | -4 | 4 | 0 | 0 | 0 | 0 | 0 | -328.8847 | -328.8911 | -0.0064 |

<p style="text-align:center">* groundstate       $\frac{\sum_{i=1}^{N}|\Delta_i|}{N} =$ 3.25E-03</p>

**Table S11.** Coefficients of each type *J* in every calculated magnetic confuguration of AgCuF$_4$ HP structure, calculated using formulas derived from Heisenberg Hamiltonian. Energies obtained from DFT calculations ($E_{DFT}$), Ising Hamiltonian ($E_H$) and differences between them ($\Delta E$, solution's error for a given state) are provided in eV. All values correspond to 2 x 2 x 2 supercell (Z=16).

| 2·S | State | $E_0$ | $J_{2D}^{Ag}$ | $J_y^{Ag}$ | $J_z^{Ag}$ | $J_x^{Ag}$ | $J_{2D}^{M}$ | $J_y^{M}$ | $J_z^{M}$ | $J_x^{M}$ | $J_1^{mix}$ | $J_2^{mix}$ | $J_3^{mix}$ | $J_4^{mix}$ | $J_5^{mix}$ | $E_{DFT}$ | $E_H$ | $\Delta E$ |
|---|---|---|---|---|---|---|---|---|---|---|---|---|---|---|---|---|---|---|
| 0 | AFM_1 | 1 | 8 | -4 | -4 | -4 | 8 | -4 | -4 | -4 | -8 | -8 | 8 | 8 | 16 | -328.2783 | -328.2756 | 0.0027 |
| 0 | AFM_2 | 1 | 8 | -4 | -4 | -4 | 8 | -4 | -4 | -4 | 8 | 8 | -8 | -8 | -16 | -328.7002 | -328.6982 | 0.0020 |
| 0 | AFM_3 | 1 | 0 | 2 | -4 | -2 | 4 | -2 | -4 | 2 | 0 | 2 | -2 | -2 | 4 | -328.5611 | -328.5607 | 0.0005 |
| 0 | AFM_4 | 1 | 0 | 4 | 4 | -4 | 8 | -4 | -4 | -4 | 0 | 0 | 0 | 0 | 0 | -328.5066 | -328.5078 | -0.0013 |
| 0 | AFM_5 | 1 | 8 | -4 | -4 | 4 | 8 | -4 | -4 | -4 | 0 | 0 | 0 | 0 | 0 | -328.4827 | -328.4846 | -0.0019 |
| 0 | AFM_6 | 1 | -8 | -4 | -4 | 4 | 8 | -4 | -4 | -4 | 0 | 0 | 0 | 0 | 0 | -328.4983 | -328.4979 | 0.0004 |
| 0 | AFM_7 | 1 | 0 | 4 | -4 | -4 | 0 | 4 | -4 | -4 | 0 | 0 | -8 | -8 | 16 | -328.3763 | -328.3771 | -0.0008 |
| 0 | AFM_8 | 1 | -8 | -4 | -4 | 4 | 0 | 4 | -4 | -4 | 0 | 0 | 0 | 0 | 0 | -328.4955 | -328.4950 | 0.0004 |
| 0 | AFM_9 | 1 | 8 | -4 | -4 | -4 | 0 | 4 | -4 | -4 | 0 | 0 | 0 | 0 | 0 | -328.4839 | -328.4840 | -0.0001 |
| 0 | AFM_10 | 1 | 0 | 4 | 4 | -4 | 0 | 4 | -4 | -4 | 0 | 0 | 0 | 0 | 0 | -328.5036 | -328.5050 | -0.0014 |
| 0 | AFM_11 | 1 | 8 | -4 | -4 | -4 | 8 | -4 | -4 | 4 | 0 | 0 | 0 | 0 | 0 | -328.4843 | -328.4838 | 0.0005 |
| 0 | AFM_12 | 1 | 8 | -4 | -4 | -4 | 8 | -4 | -4 | 4 | 0 | 0 | 0 | 0 | 0 | -328.4843 | -328.4838 | 0.0005 |
| 0 | AFM_13 | 1 | 0 | 2 | -4 | -2 | 4 | -2 | -4 | -2 | -2 | 0 | 0 | 0 | 8 | -328.5106 | -328.5094 | 0.0012 |
| 0 | AFM_14 | 1 | 0 | 4 | 4 | -4 | 8 | -4 | -4 | 4 | 0 | 0 | 0 | 0 | 0 | -328.5023 | -328.5048 | -0.0025 |
| 0 | AFM_15 | 1 | 8 | -4 | -4 | 4 | 8 | -4 | -4 | 4 | 0 | 0 | 0 | 0 | 0 | -328.4775 | -328.4815 | -0.0040 |
| 0 | AFM_16 | 1 | -8 | -4 | -4 | 4 | 8 | -4 | -4 | -4 | 0 | 0 | 0 | 0 | 0 | -328.4983 | -328.4979 | 0.0004 |
| 0 | AFM_16 | 1 | 0 | 4 | 4 | 4 | 0 | 4 | 4 | 4 | 0 | 0 | 8 | -8 | -16 | -328.3737 | -328.3716 | 0.0021 |
| 0 | AFM_17 | 1 | 0 | 3 | 3 | 3 | 0 | 3 | 3 | 3 | -2 | 2 | -6 | 6 | 12 | -328.6994 | -328.7002 | -0.0008 |
| 0 | AFM_18 | 1 | 0 | 4 | -4 | -4 | 0 | 4 | -4 | -4 | 0 | 0 | 8 | 8 | -16 | -328.6316 | -328.6320 | -0.0004 |
| **0** | **AFM_19*** | **1** | **-8** | **-4** | **-4** | **-4** | **-8** | **-4** | **-4** | **-4** | **8** | **8** | **8** | **8** | **16** | **-329.0024** | **-329.0051** | **-0.0026** |
| 8 | FM8_1 | 1 | 0 | -4 | -4 | 0 | 8 | -4 | -4 | -4 | -4 | -4 | 4 | 4 | 8 | -328.3873 | -328.3867 | 0.0006 |
| 8 | FM8_2 | 1 | 0 | -4 | -4 | 0 | 8 | -4 | -4 | -4 | 4 | 4 | -4 | -4 | -8 | -328.5981 | -328.5980 | 0.0001 |
| 8 | FM8_3 | 1 | 0 | -4 | -4 | 0 | -8 | -4 | -4 | 4 | 0 | 0 | 0 | 0 | 0 | -328.4907 | -328.4928 | -0.0022 |
| 8 | FM8_4 | 1 | 0 | -4 | -4 | 0 | 8 | -4 | -4 | 4 | 0 | 0 | 0 | 0 | 0 | -328.4888 | -328.4893 | -0.0005 |
| 16 | FM16_1 | 1 | -8 | -4 | -4 | -4 | 8 | -4 | -4 | -4 | 0 | 0 | 0 | 0 | 0 | -328.5005 | -328.5002 | 0.0003 |
| 16 | FM16_2 | 1 | -8 | -4 | -4 | -4 | -8 | -4 | -4 | 4 | 0 | 0 | 0 | 0 | 0 | -328.5026 | -328.5007 | 0.0019 |
| 16 | FM16_3 | 1 | -8 | -4 | -4 | -4 | 8 | -4 | -4 | 4 | 0 | 0 | 0 | 0 | 0 | -328.4975 | -328.4971 | 0.0004 |
| 16 | FM16_4 | 1 | -8 | -4 | -4 | -4 | 0 | 4 | -4 | -4 | 0 | 0 | 0 | 0 | 0 | -328.4992 | -328.4973 | 0.0019 |
| 16 | FM16_5 | 1 | -8 | -4 | -4 | -4 | 0 | 4 | 4 | -4 | 0 | 0 | 0 | 0 | 0 | -328.4988 | -328.5040 | -0.0051 |
| 16 | FM16_6 | 1 | -2 | 0 | 0 | 0 | -2 | 0 | 0 | 0 | -2 | -2 | -8 | 0 | 0 | -328.4150 | -328.4086 | 0.0065 |
| 16 | FM16_7 | 1 | 0 | 0 | 0 | 0 | 0 | 0 | 0 | 0 | 0 | 0 | 0 | -8 | -16 | -328.3621 | -328.3610 | 0.0011 |
| 32 | FM32 | 1 | -8 | -4 | -4 | -4 | -8 | -4 | -4 | -4 | -8 | -8 | -8 | -8 | -16 | -327.9998 | -328.0024 | -0.0026 |

<p style="text-align:center">* groundstate       $\frac{\sum_{i=1}^{N}|\Delta_i|}{N} =$ 1.56E-03</p>

**Table S12.** Coefficients of each type $J$ in every calculated magnetic confuguration of AgNiF$_4$ LP structure, calculated using formulas derived from Heisenberg Hamiltonian. Energies obtained from DFT calculations ($E_{DFT}$), Ising Hamiltonian ($E_H$) and differences between them ($\Delta E$, solution's error for a given state) are provided in eV. All values correspond to 2 x 2 x 2 supercell (Z=16).

| 2·S | State | $E_0$ | $J_{2D}^{Ag}$ | $J_y^{Ag}$ | $J_z^{Ag}$ | $J_x^{Ag}$ | $J_{2D}^{M}$ | $J_y^{M}$ | $J_z^{M}$ | $J_x^{M}$ | $J_1^{mix}$ | $J_2^{mix}$ | $J_3^{mix}$ | $J_4^{mix}$ | $J_5^{mix}$ | $E_{DFT}$ | $E_H$ | $\Delta E$ |
|---|---|---|---|---|---|---|---|---|---|---|---|---|---|---|---|---|---|---|
| 0 | AFM_1 | 1 | -8 | -4 | -4 | 4 | -32 | -16 | -16 | 16 | 0 | 0 | 0 | 0 | 0 | -374.2003 | -374.1922 | 0.0081 |
| **0** | **AFM_2*** | **1** | **8** | **-4** | **-4** | **-4** | **32** | **-16** | **-16** | **-16** | **16** | **16** | **-16** | **-16** | **-32** | **-375.1954** | **-375.1969** | **-0.0015** |
| 0 | AFM_3 | 1 | 8 | -4 | -4 | -4 | 32 | -16 | -16 | -16 | -16 | -16 | 16 | 16 | 32 | -373.9419 | -373.9417 | 0.0003 |
| 0 | AFM_4 | 1 | 8 | -4 | -4 | 4 | 32 | -16 | -16 | 16 | 0 | 0 | 0 | 0 | 0 | -374.5936 | -374.5913 | 0.0023 |
| 0 | AFM_5 | 1 | 0 | 4 | -4 | -4 | 0 | 16 | -16 | -16 | 0 | 0 | 16 | 16 | -32 | -374.2307 | -374.2290 | 0.0016 |
| 0 | AFM_6 | 1 | 0 | 4 | -4 | 4 | 0 | 16 | -16 | -16 | 0 | 0 | 0 | 0 | 0 | -374.3880 | -374.3939 | -0.0060 |
| 0 | AFM_7 | 1 | 0 | 4 | -4 | -4 | 0 | 16 | -16 | 16 | 0 | 0 | 0 | 0 | 0 | -374.3883 | -374.3832 | 0.0051 |
| 0 | AFM_8 | 1 | 0 | 4 | -4 | -4 | 0 | 16 | -16 | -16 | 0 | 0 | -16 | -16 | 32 | -374.5281 | -374.5261 | 0.0020 |
| 0 | AFM_9 | 1 | 0 | -4 | 4 | -4 | 0 | -16 | 16 | -16 | 16 | 16 | 0 | 0 | 0 | -375.0153 | -375.0121 | 0.0032 |
| 0 | AFM_10 | 1 | 0 | 4 | 4 | -4 | 0 | 16 | 16 | -16 | 0 | 0 | 0 | 0 | 0 | -374.3877 | -374.3847 | 0.0030 |
| 0 | AFM_11 | 1 | -8 | -4 | -4 | 4 | 0 | -16 | 16 | -16 | 0 | 0 | 0 | 0 | 0 | -374.3467 | -374.3497 | -0.0031 |
| 0 | AFM_12 | 1 | 0 | -4 | 4 | -4 | -32 | -16 | -16 | 16 | 0 | 0 | 0 | 0 | 0 | -374.2154 | -374.2193 | -0.0038 |
| 0 | AFM_13 | 1 | -1 | -2 | 2 | -2 | -4 | -4 | 4 | -4 | 10 | 6 | 0 | -2 | 4 | -374.7355 | -374.7324 | 0.0031 |
| 0 | AFM_14 | 1 | -1 | 2 | 2 | -2 | -4 | 4 | 4 | -4 | 2 | 0 | -2 | 0 | 4 | -374.4446 | -374.4477 | -0.0031 |
| 4 | FM4 | 1 | 0 | 2 | -4 | 2 | 0 | 8 | -16 | -8 | 4 | 0 | -4 | -4 | 8 | -374.5575 | -374.5575 | 0.0001 |
| 8 | FM8 | 1 | -4 | 0 | -4 | 0 | -32 | -16 | -16 | 16 | 0 | 0 | 0 | 0 | 0 | -374.2091 | -374.2121 | -0.0030 |
| 10 | FM10 | 1 | -1 | -1 | -1 | 1 | -16 | -16 | 0 | 0 | 6 | 6 | 4 | 2 | 8 | -374.5390 | -374.5452 | -0.0062 |
| 16 | FM16_1 | 1 | -8 | -4 | -4 | -4 | -32 | -16 | -16 | 16 | 0 | 0 | 0 | 0 | 0 | -374.1705 | -374.1758 | -0.0053 |
| 16 | FM16_2 | 1 | -8 | -4 | -4 | -4 | 0 | 16 | -16 | -16 | 0 | 0 | 0 | 0 | 0 | -374.3168 | -374.3214 | -0.0046 |
| 16 | FM16_3 | 1 | -8 | -4 | -4 | -4 | -32 | -16 | -16 | -16 | 16 | 16 | 16 | 16 | 32 | -374.8149 | -374.8131 | 0.0018 |
| 24 | FM24 | 1 | -4 | 0 | -4 | 0 | -16 | 0 | -16 | 0 | 0 | -8 | -8 | -8 | 0 | -374.2638 | -374.2646 | -0.0008 |
| 32 | FM32 | 1 | -8 | -4 | -4 | 4 | -32 | -16 | -16 | -16 | 0 | 0 | 0 | 0 | 0 | -374.1909 | -374.1866 | 0.0043 |
| 48 | FM48 | 1 | -8 | -4 | -4 | -4 | -32 | -16 | -16 | -16 | -16 | -16 | -16 | -16 | -32 | -373.5297 | -373.5272 | 0.0024 |

\* groundstate $\qquad\qquad\qquad\qquad\qquad\qquad\qquad\qquad\qquad\qquad\qquad\qquad\qquad\qquad\qquad\frac{\sum_{i=1}^{N}|\Delta_i|}{N} =$ 3.25E-03

**Table S13.** Coefficients of each type $J$ in every calculated magnetic confuguration of AgNiF$_4$ HP structure, calculated using formulas derived from Heisenberg Hamiltonian. Energies obtained from DFT calculations ($E_{DFT}$), Ising Hamiltonian ($E_H$) and differences between them ($\Delta E$, solution's error for a given state) are provided in eV. All values correspond to 2 x 2 x 2 supercell (Z=16).

| 2·S | State | $E_0$ | $J_{2D}^{Ag}$ | $J_y^{Ag}$ | $J_z^{Ag}$ | $J_x^{Ag}$ | $J_{2D}^{M}$ | $J_y^{M}$ | $J_z^{M}$ | $J_x^{M}$ | $J_1^{mix}$ | $J_2^{mix}$ | $J_3^{mix}$ | $J_4^{mix}$ | $J_5^{mix}$ | $E_{DFT}$ | $E_H$ | $\Delta E$ |
|---|---|---|---|---|---|---|---|---|---|---|---|---|---|---|---|---|---|---|
| 0 | AFM_1 | 1 | -8 | -4 | -4 | 4 | 32 | -16 | -16 | 16 | 0 | 0 | 0 | 0 | 0 | -373.4496 | -373.4435 | 0.0061 |
| 0 | AFM_2 | 1 | 8 | -4 | -4 | -4 | 32 | -16 | -16 | -16 | -16 | -16 | 16 | 16 | 32 | -373.5881 | -373.5862 | 0.0018 |
| **0** | **AFM_3** | **1** | **8** | **-4** | **-4** | **-4** | **32** | **-16** | **-16** | **-16** | **16** | **16** | **-16** | **-16** | **-32** | **-374.2088** | **-374.1987** | **0.0102** |
| 0 | AFM_4 | 1 | 8 | -4 | -4 | 4 | -32 | -16 | -16 | 16 | 0 | 0 | 0 | 0 | 0 | -373.6222 | -373.6262 | -0.0040 |
| 0 | AFM_5 | 1 | 0 | 4 | -4 | -4 | 0 | 16 | -16 | -16 | 0 | 0 | 16 | 16 | -32 | -373.6635 | -373.6636 | -0.0001 |
| 0 | AFM_6 | 1 | 0 | 4 | -4 | 4 | 0 | 16 | -16 | -16 | 0 | 0 | 0 | 0 | 0 | -373.5433 | -373.5673 | -0.0240 |
| 0 | AFM_7 | 1 | 0 | 4 | -4 | -4 | 0 | 16 | -16 | 16 | 0 | 0 | 0 | 0 | 0 | -373.5645 | -373.5569 | 0.0076 |
| 0 | AFM_8 | 1 | -8 | -4 | -4 | 4 | 0 | 16 | -16 | -16 | 0 | 0 | 0 | 0 | 0 | -373.3530 | -373.3359 | 0.0170 |
| 0 | AFM_9 | 1 | 0 | -4 | 4 | -4 | 0 | -16 | 16 | -16 | 16 | 16 | 0 | 0 | 0 | -374.0146 | -374.0311 | -0.0165 |
| 0 | AFM_10 | 1 | 0 | 4 | 4 | -4 | 0 | 16 | 16 | -16 | 0 | 0 | 0 | 0 | 0 | -373.5624 | -373.5459 | 0.0165 |
| 0 | AFM_11 | 1 | 0 | -4 | 4 | -4 | 32 | -16 | -16 | 16 | 0 | 0 | 0 | 0 | 0 | -373.6702 | -373.6696 | 0.0006 |

| | | | | | | | | | | | | | | | | | |
|---|---|---|---|---|---|---|---|---|---|---|---|---|---|---|---|---|---|
| 0 | AFM_12 | 1 | 1 | 2 | 0 | 2 | 16 | 0 | -16 | 0 | 0 | 4 | 0 | 0 | -8 | -373.6265 | -373.6277 | -0.0012 |
| 4 | FM4 | 1 | 0 | 2 | -4 | 2 | 0 | 8 | -16 | -8 | 0 | 4 | 4 | 4 | -8 | -373.5769 | -373.5757 | 0.0012 |
| 8 | FM8 | 1 | -4 | 0 | -4 | 0 | 32 | -16 | -16 | 16 | 0 | 0 | 0 | 0 | 0 | -373.5617 | -373.5616 | 0.0001 |
| 16 | FM16_1 | 1 | 0 | 0 | 0 | 0 | 0 | -16 | 0 | 0 | 4 | 4 | 4 | -4 | 8 | -373.7018 | -373.7018 | 0.0000 |
| 16 | FM16_2 | 1 | -8 | -4 | -4 | -4 | 32 | -16 | -16 | 16 | 0 | 0 | 0 | 0 | 0 | -373.4382 | -373.4483 | -0.0101 |
| 16 | FM16_3 | 1 | -8 | -4 | -4 | -4 | 0 | 16 | -16 | -16 | 0 | 0 | 0 | 0 | 0 | -373.3235 | -373.3408 | -0.0173 |
| 16 | FM16_4 | 1 | -8 | -4 | -4 | -4 | -32 | -16 | -16 | -16 | 16 | 16 | 16 | 16 | 32 | -373.9239 | -373.9162 | 0.0077 |
| 32 | FM32 | 1 | -8 | -4 | -4 | 4 | -32 | -16 | -16 | -16 | 0 | 0 | 0 | 0 | 0 | -373.2175 | -373.2126 | 0.0049 |
| 48 | FM48 | 1 | -8 | -4 | -4 | -4 | -32 | -16 | -16 | -16 | -16 | -16 | -16 | -16 | -32 | -372.5182 | -372.5187 | -0.0005 |

\* groundstate $\qquad \frac{\sum_{i=1}^{N}|\Delta_i|}{N} =$ 7.37E-03